\documentclass[journal,article,submit,pdftex,moreauthors]{Definitions/mdpi} 
\usepackage{makecell} 
\firstpage{1} 
\pubvolume{1}
\issuenum{1}
\articlenumber{0}
\pubyear{2025}
\copyrightyear{2025}
\datereceived{ } 
\daterevised{ } 
\dateaccepted{ } 
\datepublished{ } 
\hreflink{https://doi.org/} 

\usepackage{graphicx}
\usepackage{dcolumn}
\usepackage{bm}
\usepackage[normalem]{ulem}
\usepackage{xcolor}
\usepackage{float}
\usepackage{subfig}
\usepackage{slashed}
\usepackage{wrapfig}
\usepackage{amssymb}
\usepackage{amsfonts}
\usepackage{amsmath}
\usepackage{longtable}
\usepackage{rotating}
\usepackage{lscape}
\usepackage{epsfig}
\usepackage{multirow}
\usepackage{commath}
\usepackage{tabularx}
\usepackage{textcomp}
\usepackage{adjustbox}

\preto{\abstractkeywords}{\nolinenumbers}

\Title{Bayesian analysis of hybrid neutron star EOS constraints within an instantaneous nonlocal chiral quark matter model}

\TitleCitation{Bayesian analysis of hybrid neutron star EOS}

\Author{Alexander Ayriyan$^{1,2}$\orcidA{}, 
David Blaschke$^{1,3,4}$\orcidB{}, Juan Pablo Carlomagno$^{5,6}$\orcidC{}, 
Gustavo A. Contrera$^{5,6}$\orcidD{} and  Ana Gabriela Grunfeld$^{5,7}$\orcidE{}*}
\AuthorNames{Alexander Ayriyan, David Blaschke, Juan Pablo Carlomagno, Gustavo A. Contrera and Ana Gabriela Grunfeld}
\AuthorCitation{Ayriyan, A.; Blaschke, D.; Carlomagno, J.P.; Contrera, G.A.; Grunfeld, A.G.}

\AuthorNames{Alexander Ayriyan, David Blaschke, Juan Pablo Carlomagno, Gustavo A. Contrera and Ana Gabriela Grunfeld}

\isAPAStyle{%
       \AuthorCitation{Ayriyan, A., Blaschke, D., Carlomagno, J.P., Contrera, G.A., Grunfeld, A.G.}
         }{%
        \isChicagoStyle{%
        \AuthorCitation{Alexander Ayriyan, David Blaschke, Juan Pablo Carlomagno, Gustavo A. Contrera and Ana Gabriela Grunfeld.}
        }{
        \AuthorCitation{Ayriyan, A.; Blaschke, D.; Carlomagno, J.P.; Contrera, G.A.; Grunfeld, A.G.}
        }
}

\address{%
$^{1}$ \quad Institute for Theoretical Physics, University of Wroclaw, Max Born Pl. 9, 50-204, Wroclaw, Poland\\
$^{2}$ \quad A. Alikhanyan National Science Laboratory (Yerevan Physics Institute), Alikhanyan Brothers street 2, 0036 Yerevan, Armenia\\
$^{3}$ \quad Helmholtz-Zentrum Dresden-Rossendorf (HZDR), Bautzner Landstrasse 400, 01328 Dresden, Germany\\
$^{4}$ \quad Center for Advanced Systems Understanding (CASUS), Untermarkt 20, 02826 G\"orlitz, Germany\\
$^{5}$  \quad CONICET, Godoy Cruz 2290, Buenos Aires, Argentina\\
$^{6}$  \quad IFLP, UNLP, CONICET, Facultad de Ciencias Exactas, Diagonal 113 entre 63 y 64, La Plata 1900, Argentina\\
$^{7}$ \quad Departamento de F\'\i sica, Comisi\'on Nacional de Energ\'{\i}a At\'omica, Av. Libertador 8250, (1429) Buenos Aires, Argentina
}

\corres{Correspondence: alexander.ayriyan@uwr.edu.pl}

\abstract{We present a physics-informed Bayesian analysis of equation of state  constraints using observational data for masses, radii and tidal deformability of pulsars and a generic class of hybrid neutron star equation of state with color superconducting quark matter on the basis of a recently developed nonlocal chiral quark model. 
The nuclear matter phase is described within a relativistic density functional model of the DD2 class and the phase transition is obtained by a Maxwell construction. 
We find the region in the two-dimensional parameter space spanned by the vector meson coupling and the scalar diquark coupling, where three conditions are fulfilled: 1) the Maxwell construction can be performed, 2) the maximum mass of the hybrid neutron star is not smaller than 2.0 M$_\odot$ and 3) the onset density of the phase transition is not below the nuclear saturation density $n_0=0.15$ fm$^{-3}$.
The result of this study shows that the favorable neutron star equation of state has low onset masses for the occurrence of a color superconducting quark matter core between $0.5$--$0.7~M_\odot$ and maximum masses in the range $2.15$--$2.22~M_\odot$. In the typical mass range of $1.2$--$2.0~M_\odot$, the radii of these stars are between 11.9 and 12.4 km, almost independent of the mass. In  principle, hybrid stars would allow for larger maximum masses than provided by the hadronic reference equation of state.}


\begin{document}

\section{Introduction}

Exploring the phases of matter under extreme conditions, like those in 
heavy-ion collisions \cite{Sorensen:2023zkk} or within neutron stars
\cite{Blaschke:2016}, is a topic that has attracted significant attention over the past decades \cite{Rezzolla:2018jee}. 
We are particularly interested in studying transitions between different phases of strongly interacting matter, like quark-gluon plasma, color superconducting quark matter, and hadronic matter. 
These transitions have a big impact on astrophysical phenomena like 
the formation of eccentric binaries and isolated millisecond pulsars (MSPs) \cite{Chanlaridis:2024rov}, binary neutron star mergers \cite{Bauswein:2018bma}, 
and even trigger supernova explosions of massive blue supergiant stars \cite{Fischer:2017lag} 
to mention just a few recent applications.

The status of quark matter as a state of matter in the inner core of neutron stars is still controversial. Recall its prehistory.
Immediately after the introduction of the quark model in 1964 by Gell-Mann \cite{Gell-Mann:1964ewy} and Zweig \cite{Zweig:1964ruk}, Ivanenko and Kurdgelaidze proposed the existence of quark stars \cite{Ivanenko:1965dg}. It was as early as 1968 that Gerlach suggested a gravitational instability due to a strong phase transition, which would entail that neutron stars with quark matter cores form a separate ``third family'' of compact stars \cite{Gerlach:1968zz}.
When quantum chromodynamics (QCD) was conceived as a nonabelian gauge field theory of strong interactions with quarks and gluons as elementary degrees of freedom with the property that their interactions weaken at high energies and momentum transfers (asymptotic freedom \cite{Gross:1973id,Politzer:1973fx}), the conjecture was prompted that superdense matter in neutron star cores shall consist of quarks rather than neutrons \cite{Collins:1974ky}. 
While this was merely a qualitative statement about the possibility of quark deconfinement in neutron star interiors, a quantitative estimate has been performed by Baym and Chin \cite{Baym:1976yu} on the basis of the M.I.T. bag model \cite{Chodos:1974je,Chodos:1974pn,DeGrand:1975cf} and the by then typical nuclear matter equations of state (EOS), the relativistic Walecka mean field model \cite{Walecka:1974qa,Chin:1974sa} and the variational approach by Pandharipande and Smith \cite{Pandharipande:1975zev} exploiting the Reid potential. The transition densities obtained by Baym and Chin were beyond tenfold nuclear matter density and thus too large for quark deconfinement to occur in neutron stars \cite{Baym:1976yu}.
Baym and Chin argued against the existence of Gerlach's third family stars.
Considering the absolutely stable strange quark matter hypothesis \cite{Bodmer:1971we,Witten:1984rs}, Haensel et al. \cite{Haensel:1986qb} came to the conclusion that strange quark stars with a thin nuclear crust may exist because they found an early transition in very light stars of less than half a solar mass, even as a third family.  
However, immediately after that, Bethe et al. \cite{Bethe:1987sv} reconsidered the discussion by including running coupling of quarks to the bag model EOS and concluded that the transition density was too high to occur in neutron stars.

The discussion of ``pro'' and ``con'' quark matter in compact stars has continued ever since and keeps being fueled by new observational data on masses and radii of pulsars, but also of features revealed in the observation of neutron star mergers as GW170817 \cite{LIGOScientific:2017vwq,LIGOScientific:2018cki} and its kilonova \cite{LIGOScientific:2017ync,Sneppen:2024jch}. 
For recent comprehensive analyses of hybrid EOS constraints in this context, see \cite{Blacker:2024tet,Christian:2025dhe} and references therein.
However, since the available data on masses and radii of pulsars do not allow for a firm model-independent conclusion about a quark deconfinement transition, it has been tried to apply agnostic Bayesian analysis methods to check the evidence for nontrivial phase-transition related features 
\cite{Mroczek:2023zxo}, in particular a strong phase transition \cite{Brandes:2023hma}.
In their study, the authors of \cite{Brandes:2023hma} found evidence against a strong first-order (deconfinement) phase transition in neutron stars.
However, since the lightest neutron star to be formed as a result of a supernova explosion according to state-of-the-art simulations has a mass of at least $1.1~M_\odot$, and there is no scenario available that would explain the existence of a lighter neutron star, the recent measurement of a low mass and radius for HESS J1731-347 \cite{Doroshenko2022-uq} has been considered very skeptical in the community. Also because the authors themselves could not present an idea about the origin of a neutron star in the subsolar mass region. 
Therefore, direct evidence for a strong phase transition from mass and radius measurements in the subsolar mass range is presently lacking the observational basis. 

In this situation, we want to suggest a physics-informed Bayesian analysis
of modern multi-messenger mass and radius measurements along the lines of our earlier work \cite{Ayriyan:2021prr}, which can now be made with the new, presently available set of observational data.
Such a physics-informed study will be based on a class of hybrid equations of state (EOS) within the two-phase formalism that uses separate models for hadronic and quark matter phases joined by a phase transition construction. 
Given an appropriate hadronic EOS and the choice of the phase transition construction, there are two free parameters of the color superconducting quark matter model to be determined which cannot be reliably fixed with vacuum properties of hadrons. These are the couplings of the vector meson ($G_V$) and diquark ($G_D$) currents to the corresponding mean fields.
As has been shown in \cite{Klahn:2006iw} using the three-flavor color superconducting Nambu--Jona-Lasinio (NJL) model \cite{Blaschke:2005uj}, that with such a setting the required maximum mass $M_{\rm max}\sim 2~M_\odot$  can be achieved by a sufficiently large $G_V$, while an early onset of deconfinement in the range of typical or even subsolar neutron star masses is achieved by a large value of $G_D$. 
In a first systematic study employing this setting, in Ref.~\cite{Klahn:2013kga}, the regions in the two-dimensional plane of these two coupling strengths have been identified, where 
no stable hybrid stars were possible and where the onset of deconfinement is below a certain density. In between these two limits, the lines were shown along which the constraint on the minimal value of the maximum mass could be fulfilled.
In these early studies the Dirac-Brueckner-Hartree-Fock (DBHF) hadronic EOS \cite{Fuchs:2003zn} was used together with a Maxwell construction for the phase transition. 
An alternative study by Baym et al. \cite{Baym:2019iky} came to similar conclusions for the preferable region of parameter values in the plane of vector meson and diquark coupling constants using, however, a softer APR-like hadronic EOS and a crossover construction \cite{Masuda:2012ed}, see also the discussion in the review \cite{Baym:2017whm}.
Studies of quark matter based on the NJL model are limited due to the model's lack of confinement.
This feature becomes particularly problematic in applications to the range of finite temperatures, as in supernova explosions and binary neutron star mergers, where it would predict deconfinement at a too low temperature of about $50$--$70$ MeV.

In order to overcome this problem, two main routes have been suggested: (A) a confining density functional approach \cite{Kaltenborn:2017hus} as a relativistic formulation of the string-flip model \cite{Horowitz:1985tx,Ropke:1986qs} which was recently generalized to include diquark condensation and chiral symmetry \cite{Ivanytskyi:2022oxv,Ivanytskyi:2022wln}; and (B) density-dependent quark mass models 
(see, e.g., \cite{Fowler:1981rp,Plumer:1984aw,Wen:2005uf,Yin:2008me,Xia:2014zaa,Li:2015ida,Lugones:2022upj,Issifu:2023qoo}), in which the bag pressure is interpreted as a density-dependent scalar mean field that can be reinterpreted as a contribution to the dynamical quark mass.
The confining models of class (A) can be mapped by a Taylor expansion of the confining density functional w.r.t. the underlying quark bilinears up to second order onto an effective
NJL model with medium dependent masses. Equivalently, one can also fit the pressure as a thermodynamic potential of the confining density functional approach by adding a medium dependent bag pressure and vector meson mean field to the nonlocal NJL model \cite{Alvarez-Castillo:2018pve,Contrera:2022tqh}.
In the present work, we will employ the latter approach to a confining quark matter EOS as the basis for the physics-informed Bayesian analysis to be performed.

\section{Hybrid neutron star EOS and fitting results}

There are two main strategies to construct hybrid neutron star EOS which would fulfill the maximum mass constraint that requires the most massive hybrid star configuration to have at least $2~M_\odot$ at a radius around $12$ km.  
One can start with a nuclear matter EOS that is relatively soft at supranuclear densities, like the well-known variational EOS by Akmal, Pandharipande and Ravenhall (APR) \cite{Akmal:1998cf} or its updated versions by Togashi and coworkers \cite{Togashi:2013bfg,Togashi:2016fky,Togashi:2017mjp}.
Then, a crossover construction like the one by Masuda et al. \cite{Masuda:2012ed} is required to facilitate the transition to a stiffer quark matter EOS at high densities in a thermodynamically consistent way.
This strategy has been applied, e.g., by Baym et al. \cite{Baym:2019iky} and by Ayriyan et al. \cite{Ayriyan:2021prr}, who realized the two-zone interpolation scheme (TZIS) that was described in \cite{Baym:2017whm}. See also \cite{Ivanytskyi:2022wln} for the extension of the TZIS to finite temperature and multiple critical endpoints.

The alternative is to start from a stiff nuclear EOS, for which a Maxwell construction to a sufficiently stiff quark matter EOS can be performed, so that despite a softening of the EOS due to the phase transition nevertheless the maximum mass constraint can be fulfilled. 
As examples for such stiff hadronic EOS we mention the DBHF \cite{Fuchs:2003zn} which was used in \cite{Klahn:2006iw,Klahn:2013kga} and the relativistic density functional EOS with density-dependent nucleon-meson couplings, denoted as DD2 \cite{Typel:2009sy} which was recently used, e.g., in \cite{Ivanytskyi:2022oxv,Ivanytskyi:2022wln,Contrera:2022tqh,Carlomagno:2023nrc,Gartlein:2023vif,Gartlein:2024cbj}. This latter strategy will be applied in the present work.

\subsection{Nuclear matter phase}

For the nuclear matter phase that extends from the core\footnote{In the case of a hybrid star configuration, this means the outer core being separated  by the deconfinement transition from the inner core that is constituted by phases of deconfined quark matter.} to the crust, we employ the relativistic density functional EOS with density-dependent couplings of the nucleons to the mesonic mean fields (DD2) for which the parametrization is described in Ref. \cite{Typel:2009sy}. 
This EOS describes well all known nuclear matter properties including the nuclear symmetry energy and its slope at saturation density 
$n_{\rm sat}=0.15$ fm$^{-3}$, as well as the properties of finite nuclei that make their appearance at subnuclear densities defining the crust-core boundary. While in the inner crust these nuclei are still immersed in the Fermi seas of neutrons and electrons (A-e-n phase), the neutron drip density of about $10^{-4}$ fm$^{-3}$ marks the transition to the outer crust made up of nuclei and electrons only (A-e phase).

It is customary to employ a specific EOS for the descripton of the crust of neutron stars, like the one by Baym, Pethick and Sutherland (BPS) \cite{Baym:1971pw}, and to match it with the EOS of homogeneous dense nuclear matter phases at the crust-core transition, see \citet{Fortin:2016hny}.
In the present work, however, we break with this transition by using the EOS table of the generalized RDF (GRDF) which gives a consistent description based on the DD2 interaction of nuclear matter in the density range from $10^{-9}$ fm$^{-3}$ to $1.0$ fm$^{-3}$, i.e. from the low-density phases with nuclear clusters to homogeneous nuclear matter, including cluster dissociation. This is described more in detail in the reviews by \citet{Oertel:2016bki} and \citet{Typel:2018wmm}.
This GRDF EOS which is denoted ``DD2'' throughout this work can be retrieved from the CompOSE repository \cite{Antonopoulou:2022yot} using the manual \cite{CompOSECoreTeam:2022ddl}.

\subsection{Color superconducting 2SC quark matter}

In this work, we consider the scenario of two-flavor color-superconducting quark matter appearing in the cores of NSs at sufficiently high densities.
For its description, we employ the 3DFF nonlocal chiral quark model \cite{Contrera:2022tqh} defined by an effective Euclidean action functional that in the case of two light flavors is given by
\begin{eqnarray}
S_E &=& \int d^4 x \ \left\{ \bar \psi (x) \left(- i \rlap/\partial + \hat{m}
- \gamma_0 \hat{\mu} \right) \psi (x) - \frac{G_S}{2}\left[  j^f_S(x) j^f_S(x) \right.\right.
\nonumber \\
&+& \left.\left. \eta_D \left[j^a_D(x)\right]^\dagger j^a_D(x) 
{-} \eta_V j_V^{\mu}(x)\, j_V^{{\mu}}(x) \right] \right\} .
\label{action}
\end{eqnarray}

\noindent
Here $\hat{m}=\mathrm{diag}(m_u,m_d)$ is the current quark mass matrix with the masses for $u$ and $d$ quarks, which are assumed to coincide $m_u=m_d=2.3$ MeV, whereas $\hat{\mu}=\mathrm{diag}(\mu_u,\mu_d)$ is the diagonal matrix for the chemical potentials of $u$ and $d$ quarks, which are to be adjusted so as to fulfill charge neutrality and $\beta$-equilibrium constraints relevant for neutron  star matter. As a result remains the baryon chemical potential $\mu_B=3\mu=3(\mu_u+\mu_d)/2$ as independent thermodynamic variable.
The coupling strength in the scalar meson channel is $G_S=9.92$ GeV$^{-2}$ and $\eta_V=G_V/G_S$ ($\eta_D=G_D/G_S$ ) is the relative coupling in the vector meson (scalar diquark) channel. 
The currents are given by nonlocal operators based on a separable approximation to the effective one gluon exchange (OGE) model of QCD~\cite{Blaschke:2007ri, GomezDumm:2005hy}, 
\begin{eqnarray}
j^f_S (x) &=& \int d^4 z \  g_S(z) \ \bar \psi(x+\frac{z}{2}) \ \Gamma_f\,
\psi(x-\frac{z}{2})\,,
\nonumber 
\\
j^a_D (x) &=&  \int d^4 z \ g_D(z)\ \bar \psi_C(x+\frac{z}{2}) \ i
\gamma_5 \tau_2 \lambda_a \ \psi(x-\frac{z}{2})\,, 
\nonumber
\\
{j^\mu_V (x)} &=& \int d^4 z \ g_V(z)\ \bar \psi(x+\frac{z}{2})~{i}\gamma^\mu
\ \psi(x-\frac{z}{2})\,,
\label{cuOGE}
\end{eqnarray}
where we defined $\psi_C(x) = \gamma_2\gamma_4 \,\bar \psi^T(x)$ and $\Gamma_f=(\mathbf{1},i\gamma_5\vec\tau)$, while $\vec \tau$ and $\lambda_a$, with $a=2,5,7$, stand for Pauli and Gell-Mann matrices acting on flavor and color spaces, respectively.
The functions $g_i(z)$, $i=S,D,V$, in Eqs.~(\ref{cuOGE}) are nonlocal ``instantaneous'' form factors (3D-FF) characterizing the effective quark interaction, which depend on the spatial components of the position 4-vector $z$ only. For the nonlocality a Gaussian ansatz is employed which after Fourier transformation to the momentum space reads
\begin{equation}
    g_i(\vec{p})=\exp(-\vec{p}^2/\Lambda_i^2),~~i=S,D,
\end{equation}
with the 3-momentum vector $\vec{p}$ and the effective range 
$\Lambda_S=\Lambda_D=885.47$ MeV fitted to Coulomb-gauge lattice QCD data \cite{Burgio:2012ph}. 
Note that following \cite{Contrera:2022tqh}, in the present work we consider the vector current in Eq.~(\ref{cuOGE}) to be local, $g_V(z)=\delta(z)$.
This restriction could be easily relaxed, see \cite{Ivanytskyi:2024zip}.
The equation of state $p(T,\mu)=T\ln Z(T,\mu)$ for this model follows from the path-integral representation of the partition function $Z(T,\mu)$ which is evaluated at the mean-field level and results in the pressure 
$p_{MF}(\mu_B)$ which, following \cite{Contrera:2022tqh}, is augmented by a $\mu_B$-dependent bag pressure as in Ref. \cite{Alvarez-Castillo:2018pve}, given by the equation 
\begin{equation}
B(\mu_B) = B_0 \, f_< (\mu_B)    
\end{equation}
with
\begin{equation}
f_< (\mu_B) = \frac{1}{2}\left[1 - \mathrm{tanh} \left( \frac{\mu_B - \mu_<}{\Gamma_<} \right)\right],  
\end{equation}
where we use $\mu_< = 895$ MeV, $\Gamma_< = 180$ MeV and $B_0 = 35$ MeV/fm$^3$ as the optimal values to reproduce the astrophysical constraints from multi-messenger astronomy of pulsars.
In the present work we are interested in describing the hybrid EOS for cold compact stellar systems. Therefore, we will take the zero-temperature limit.
The resulting zero-temperature pressure $p(\mu_B)=p_{MF}(\mu_B)-B(\mu_B)$ 
leaves the two couplings $\eta_V$ and $\eta_D$ as free parameters of the 3DFF color superconducting quark matter model EOS which forms the basis of the present work. Each EOS model of the present study can be characterized as a point in the two-dimensional parameter space spanned by the pair ($\eta_V$,$\eta_D$) which will therefore play a central role. 

In order to perform a Bayesian Analysis, we fit the above introduced quark matter EOS to the constant speed of sound (CSS) form of the EOS 
\cite{Shahrbaf:2021cjz}
%
\begin{eqnarray}
\label{eq:css}
p=A \left(\frac{\mu}{\mu_x} \right)^{1+1/c_s^2} - B
\end{eqnarray}
where the $\mu=\mu_B/3$ is the quark chemical potential and $\mu_x=1$ GeV sets a scale.
Before, such a fit has been provided in \cite{Shahrbaf:2021cjz} for the covariant nonlocal chiral quark model EOS. 
Here, we present the fit (\ref{eq:css}) of the 3DFF nonlocal, color superconducting quark matter model \cite{Contrera:2022tqh} in the form 
%
\begin{eqnarray}
\label{eq:A}    
A\left(\eta_{D}, \eta_{V}\right)&=&a_1\eta_{D}^2+b_1\eta_{V}^2+c_1\eta_{D}\eta_{V}+d_1\eta_{D}+e_1\eta_{V}+f_1,\\
\label{eq:Cs}
B\left(\eta_{D}, \eta_{V}\right)&=&a_2\eta_{D}^2+b_2\eta_{V}^2+c_2\eta_{D}\eta_{V}+d_2\eta_{D}+e_2\eta_{V}+f_2,\\
\label{eq:B}  
c_s^2\left(\eta_{D}, \eta_{V}\right)&=&a_3\eta_{D}^2+b_3\eta_{V}^2+c_3\eta_{D}\eta_{V}+d_3\eta_{D}+e_3\eta_{V}+f_3,
\end{eqnarray}
The fitted parameter values are given in Tab. \ref{tab:param}. The last column presents the mean squared error (MSE) and the sum of relative squared errors (RSE).

\begin{table}[H]
\caption{Values of the parameters of Eqs. (\ref{eq:A}) - (\ref{eq:B}).}
\centering
\addtolength{\tabcolsep}{-2pt}
\renewcommand{\arraystretch}{1.1}
\begin{tabular}{c|c|c|r|r|r|r|r|r|p{1.5cm}@{}} 
\specialrule{0.1em}{0em}{0em}
$i$ & par. & units & $a_i\;\quad$ & $b_i\;\quad$ & $c_i\;\quad$ & $d_i\;\quad$ & $e_i\;\quad$ & 
$f_i\;\quad$ & \begin{tabular}{c} \phantom{0}MSE\phantom{0} \\ \hline RSE\end{tabular} \\ 
\hline
1 & $A$ & $\dfrac{\mathrm{MeV}}{\mathrm{fm}^3}$ & 88.637 & 1.861 & $-$30.368 & $-$76.654 & 2.891 & 95.324 &
\begin{tabular}{c} 3.23e{$-$3} \\ \hline 6.87e{$-$5}\end{tabular} \\ 
\hline
2 & $B$ & $\dfrac{\mathrm{MeV}}{\mathrm{fm}^3}$ & $-$192.28\phantom{0} & $-$10.180 & 30.455 & 390.92\phantom{0} & $-$28.613 & $-$100.95\phantom{0} & \begin{tabular}{c} 5.33e{$-$2} \\ \hline 1.51e{$-$3}\end{tabular} \\ 
\hline
3 & $c_s^2$ & $c^2$ & 0.142 & $-$0.071 & 0.066 & $-$0.166 & 0.134 & 0.427 & \begin{tabular}{c} 7.76e{$-$8} \\ \hline 6.42e{$-$5}\end{tabular} \\
\specialrule{0.1em}{0em}{0em}
\end{tabular}
\label{tab:param}
\end{table}

\begin{figure}[!htb]
    \centering
    \includegraphics[width=0.49\textwidth]{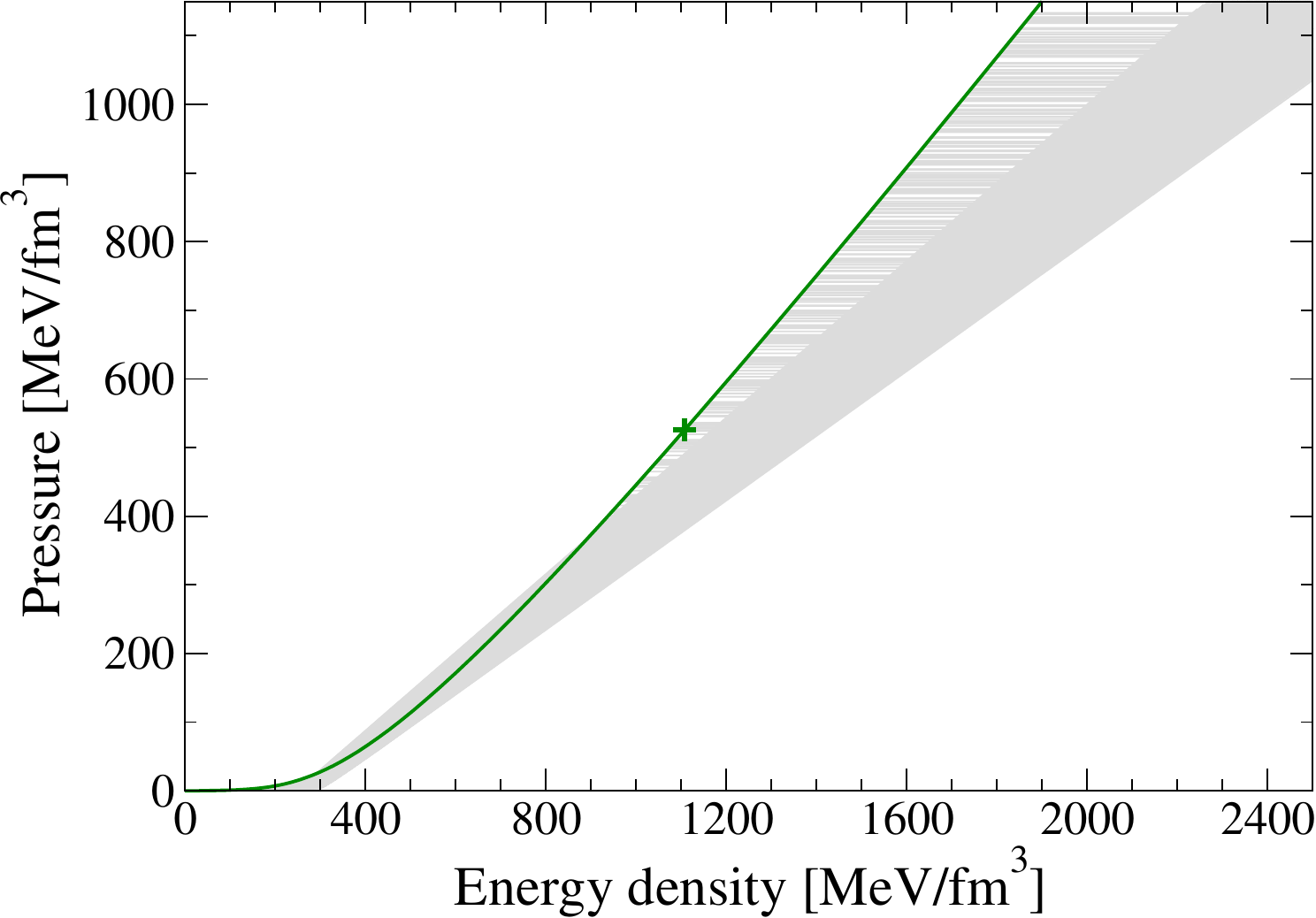}
    \hfill
    \includegraphics[width=0.49\textwidth]{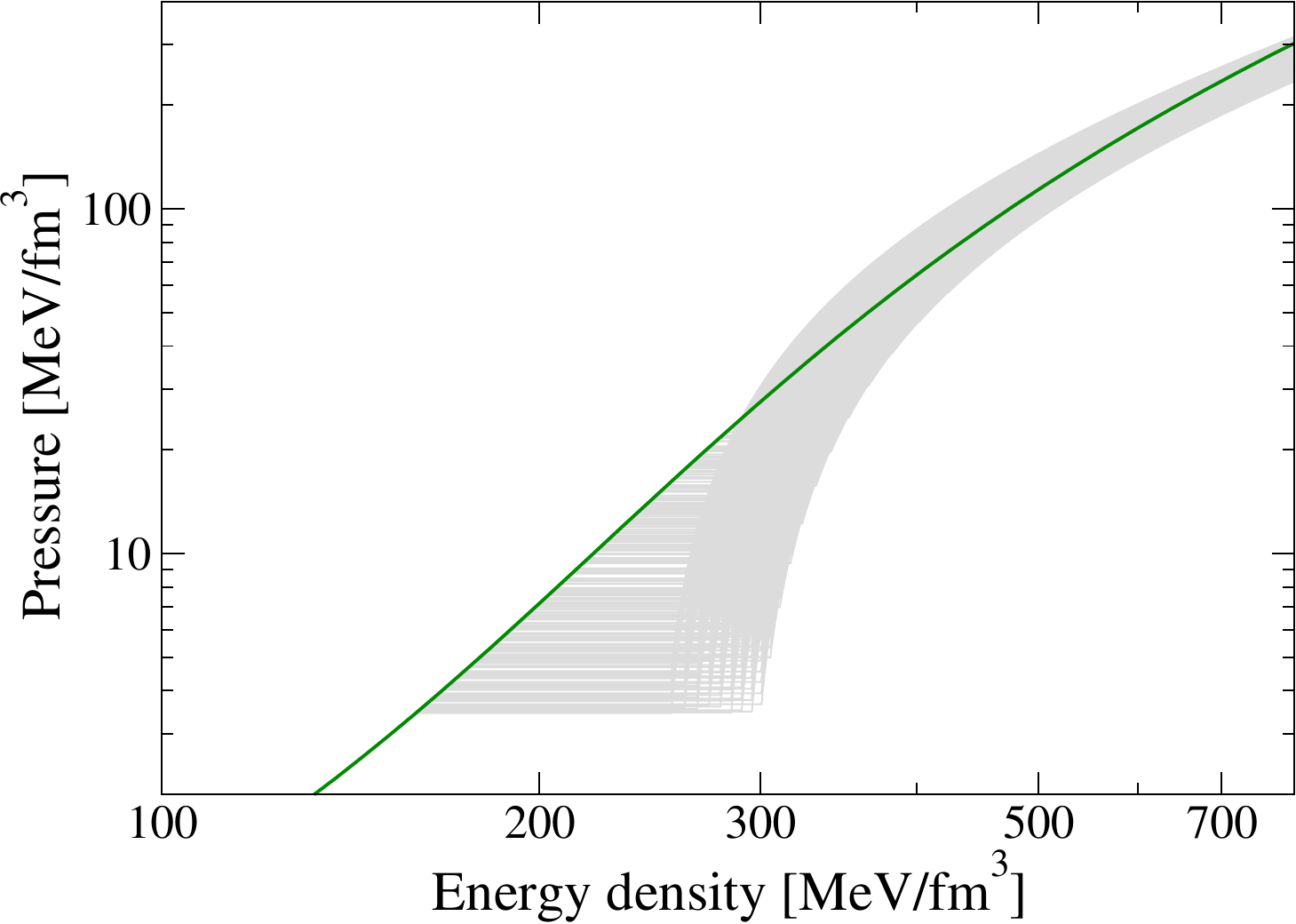}
    \caption{Dependence of pressure on energy density, which shows the jump in energy density at the phase transition from  hadronic DD2 EOS (green solid line) to the color superconducting quark matter EOS in its CSS parametrization (left panel). The right panel shows a closeup of the phase transition region at lower energy densities than that for which the maximum mass configuration of the hadronic DD2 EOS is obtained (indicated by a plus sign on the left panel).}
\label{fig:p-e}
\end{figure}

\subsection{Phase transition construction}

We employ the Maxwell construction of the nuclear-to-quark matter phase transition. The resulting EOS is characterized by a critical pressure $P_c=P_h(\mu_c)=P_q(\mu_c)$ for the onset of the transition from the hadronic phase $P_h(\mu)$ for $\mu\le \mu_c$ to the 2SC quark matter phase $P_q(\mu)$ for $\mu\ge \mu_c$ with a jump in energy density
$\Delta \varepsilon=\varepsilon_q(\mu_c)-\varepsilon_h(\mu_c)$.
In Fig. \ref{fig:p-e}, we show the hybrid neutron star EOS after Maxwell construction has been performed. In the left panel, we show the whole range of energy densities considered here, while in the right panel a zoom into the region relevant for the phase transition is shown.

\begin{figure}[htb]
    \centering
    \includegraphics[width=0.8\textwidth]{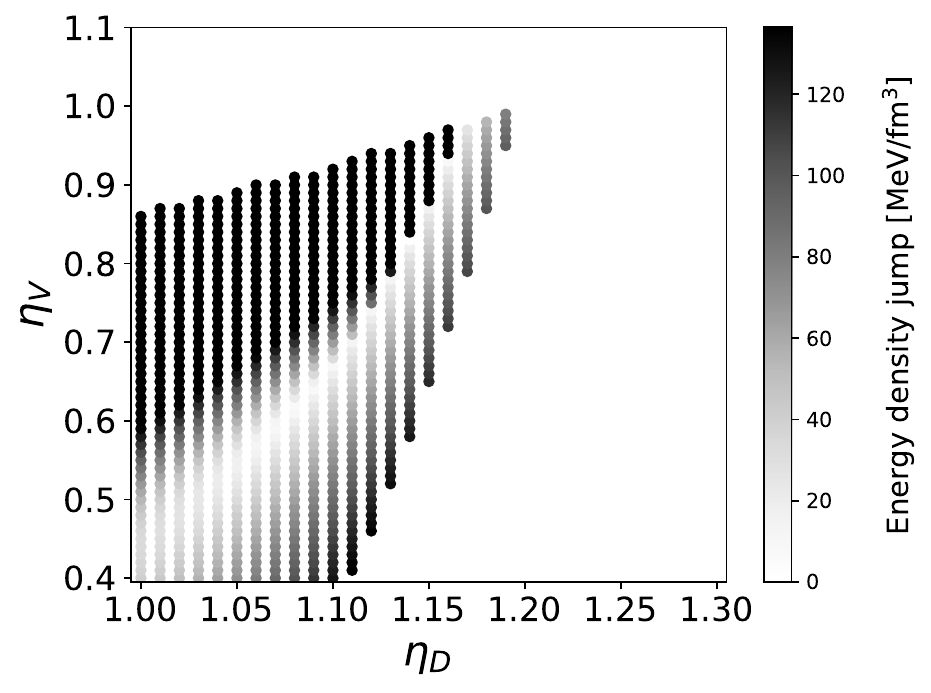}
    \caption{Diagram showing the parameter region in the plane of the coupling constants $\eta_D$--$\eta_V$, where a Maxwell construction is possible. 
    The darkness of the grayscale dots corresponds to the size of the jump in energy density $\Delta \varepsilon$ at the phase transition from the hadronic DD2 to the color superconducting 3DFF quark matter EOS. 
    The white spots in the middle of the highlighted region stand for parameter sets for which the transition degenerates to a crossover with $\Delta \varepsilon =0$.}
\label{fig:eDeV}
\end{figure}

For some values of the parameters $(\eta_D,\eta_V)$ there is no crossing of the hadronic and quark-matter pressure curves so that no Maxwell construction can be performed. 
In Fig. \ref{fig:eDeV}, we show the region in the $\eta_D$--$\eta_V$ plane where a Maxwell construction is possible by attributing the darkness of the grayscale to the size of the jump in the energy density $\Delta \varepsilon=\varepsilon_q(\mu_c)-\varepsilon_h(\mu_c)$.

\section{Mass, radius and tidal deformability fom TOV solutions}

There is a one-to-one relationship between the observable mass-radius relation for neutron stars and the EOS of neutron star matter \cite{1992ApJ398569L} which follows from the Tolman-Oppenheimer-Volkoff (TOV) equations \cite{Tolman:1939jz,Oppenheimer:1939ne} of general relativistic hydrodynamic stability of spherically symmetric, nonrotating compact stars. These equations form the basis for extracting information about the EOS from observations.
In the following we will present those mass and radius measurements which provide at present the most promising EOS constraints. In addition, due to the measurement of the gravitational wave signal from the inspiral phase of the binary neutron star merger event GW170817 by the LIGO-Virgo Collaboration \cite{LIGOScientific:2018cki}, there is a constraint on the tidal deformability parameter of neutron stars, see \cite{Hinderer:2007mb,Damour:2009vw,Hinderer:2009ca} for the system of differential equations. 

\begin{figure}[!thb]
    \centering
    \includegraphics[width=0.75\linewidth]{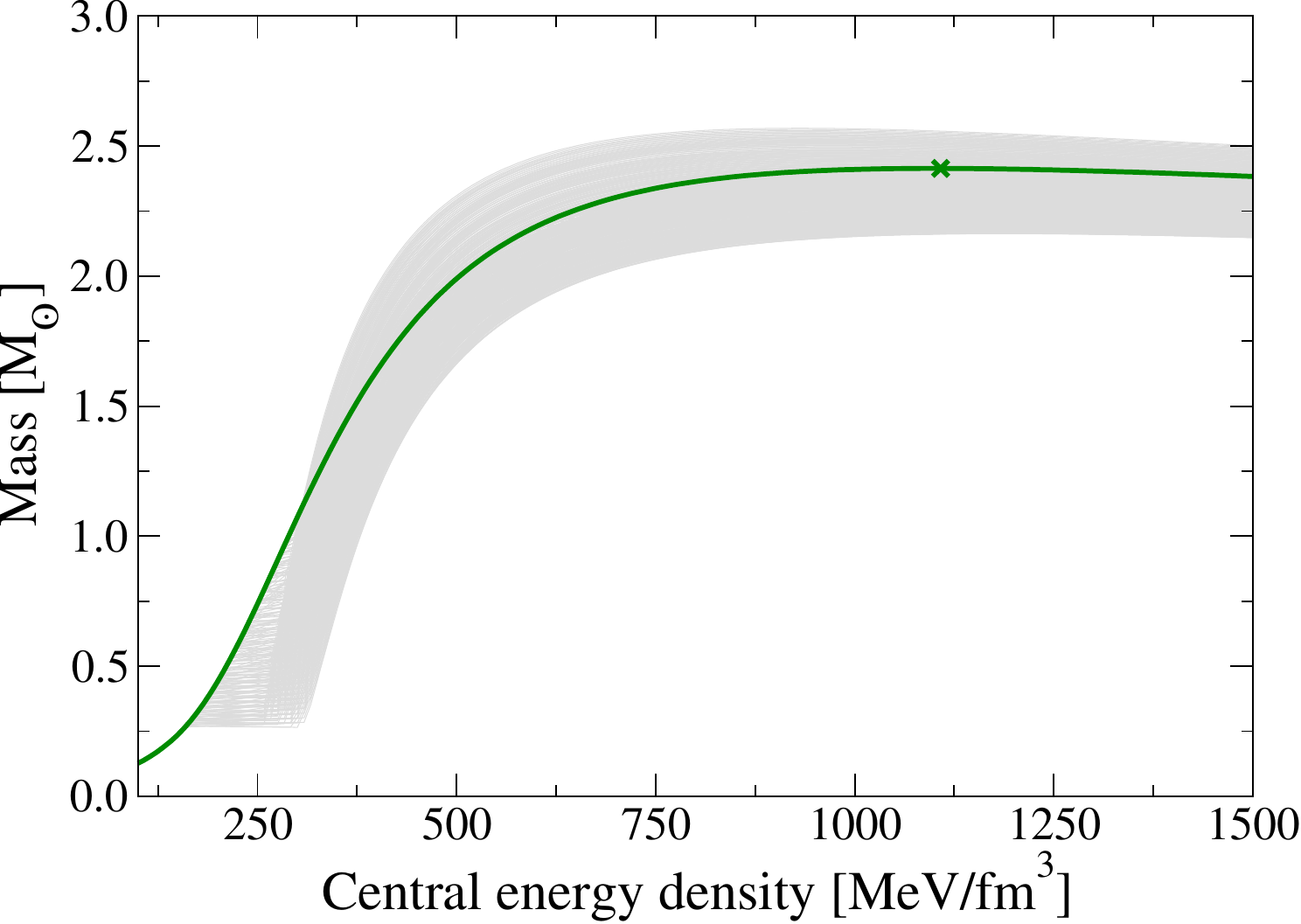}
    \caption{
    Mass versus central energy density
for the hybrid EOS with a phase transition from the hadronic DD2 EOS (green line) to color superconducting quark matter described in the nonlocal 3DFF model for varying EOS parameters $\eta_V$ and $,\eta_D$ (grey lines). The maximum mass configuration among all purely baryonic ones obtained with the DD2 EOS is denoted with a cross on the green line.}
\label{fig:Meps}
\end{figure}

In Fig. \ref{fig:Meps} we show the dependence of the neutron star mass on its central energy density for the EOS parameterizations provided in the previous section.
We note that there are cases of hybrid star sequences which have a maximum mass exceeding that of the hadronic DD2 reference EOS shown by the green solid line. On that figure we indicate by a green ``X'' the location of the maximum mass configuration of the purely hadronic case which has $M_{\rm max}=2.414~M_\odot$ and a central energy density $\varepsilon_c^{\rm max}=1107.8$ MeV/fm$^3$.

\begin{figure}[!htb]
    \centering
    \includegraphics[width=0.77\textwidth]{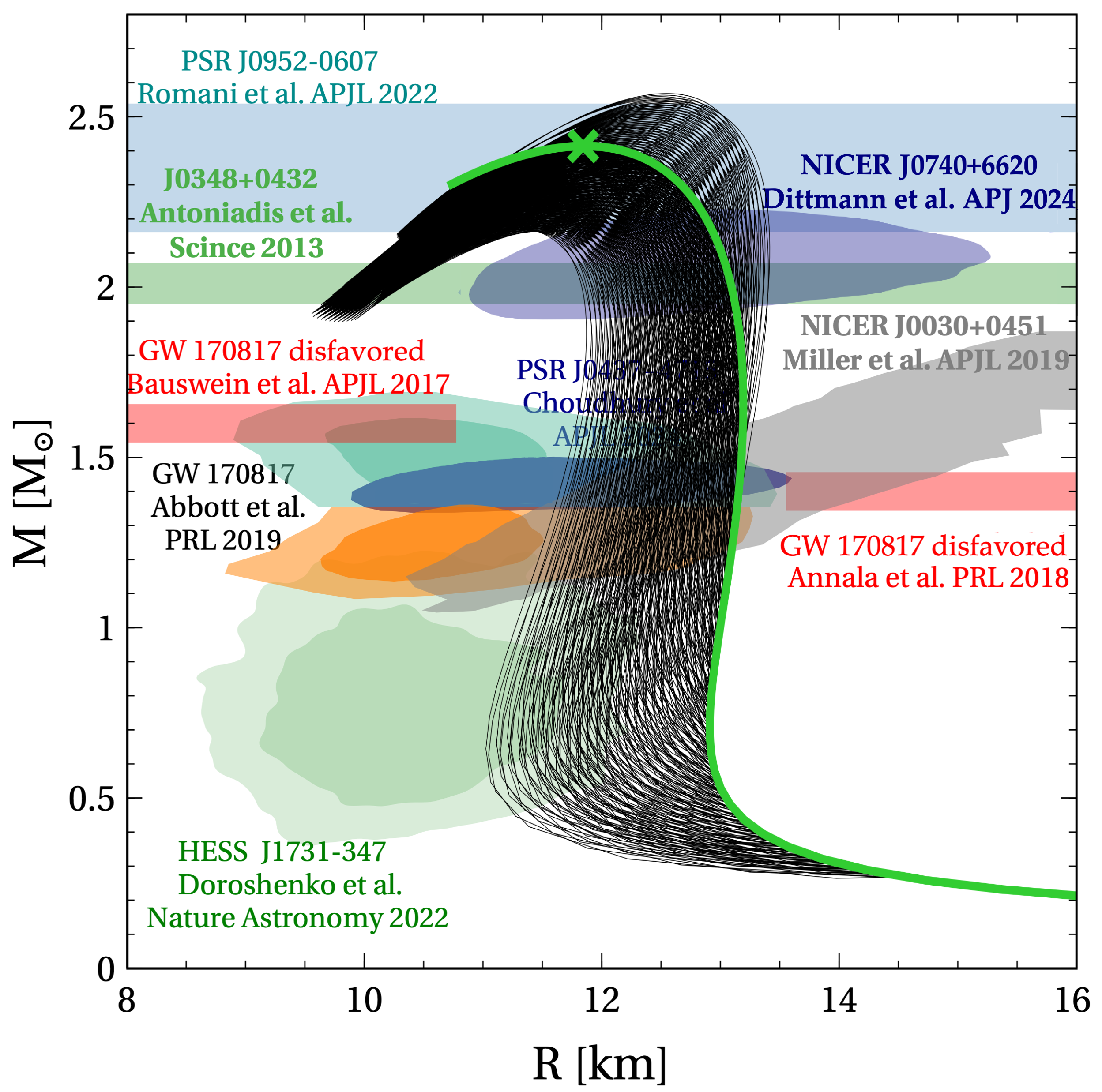}
    \caption{Same as Fig. \ref{fig:Meps} for the mass-radius relations as solutions of the TOV equations (black solid lines). Colored regions indicate observational constraints that are discussed in the text.}
\label{fig:MR}
\end{figure}

In Fig. \ref{fig:MR} we show as thin black lines the mass-radius relationships that are obtained as solutions of the TOV equations for the set of hybrid EOS introduced in the previous section. For a comparison we show also here as a green solid line the purely hadronic sequence of neutron star configurations that follows for the hadronic baseline EOS DD2. The maximum mass configuration is indicated by a green cross\footnote{We would like to draw the attention of the reader to the appearance of Moir\'e-like structures in the set of lines in the $M(R)$ diagram of Fig. \ref{fig:MR} that resemble sequences of hybrid stars generated by an equidistant variation of the quark matter EOS parameters $\eta_D$ und $\eta_S$. 
These structures originate from crossings of the $M(R)$ lines that collimate in the so-called special points \cite{Yudin:2014mla,Cierniak:2020eyh} that are lined up along the ``trains'' \cite{Shahrbaf:2021cjz,Blaschke:2022gql}.}. 
We note that some of the hybrid star sequences (shown by thin black lines) have maximum masses exceeding the value for the purely hadronic neutron star case. This effect apparently contradicts the naive expectation that in the case of a first-order phase transition, the maximum mass should generally be lowered because of the related softening of the EOS. 
While this effect is observed here too, it may get reversed when the quark-matter EOS at high densities becomes stiffer than the hadronic one. 

Speaking about an EOS that stiffens at high densities prompts the question whether the limiting case of perturbative QCD (pQCD) at about 40 times the saturation density can be reached from the highest densities that are met in the center of the maximum mass configurations of our (hybrid) EOS without violating the minimal constraints of causality ($c_s^2= dP/d\varepsilon \le 1$) and thermodynamic stability ($dP/d\varepsilon \ge 0$).
This question has been first considered by Komoltsev and Kurkela in \cite{Komoltsev:2021jzg} and was subsequently extended in Refs. \cite{Gorda:2022jvk,Annala:2023cwx} with the result that some of the neutron star EOS from the CompOSE repository \cite{Typel:2021aqw,Antonopoulou:2022yot,CompOSECoreTeam:2022ddl} would not meet this pQCD constraint. 
We tested our set of (hybrid) EOS using the program provided by \cite{Komoltsev:2023} and found that the entire set meets the pQCD constraint. 

For comparison, we show in Fig. \ref{fig:MR} also the regions of mass and radius measurements
that will be used to constrain the EOS within the Bayesian analysis performed in this work.
The set of data selected for the present study is similar to the choice made by \citet{Brandes:2023hma}. 
It consists of a basic set of observational constraints and additional ones that may have a disputed status, or it may be interesting to study the effect of their inclusion on the results of the Bayesian analysis.
The basic set contains the precise mass measurement of the high-mass pulsar PSR J0348+0432 in binary form with a white dwarf.
It provides a lower limit for the maximum mass:  
\begin{itemize}
    \item[{(I)}] $M = 2.01_{-0.04}^{+0.04}~M_{\odot}$ for PSR J0348+0432 by~\citet{Antoniadis:Science:2013} \footnote{It is worth to note that a recent estimation published on arXiv \cite{Saffer:arxiv:2024}, incorporating data from the Canadian Hydrogen Intensity Mapping Experiment \cite{CHIME2022}, suggests a lower mass, $M = 1.806_{-0.037}^{+0.037}~M_{\odot}$. However, there are indications that there is a third body present in this system so that the analysis has to be reconsidered \cite{Antoniadis:2025}.}
\end{itemize}
\noindent
In addition, these simultaneous mass-radius measurements of the NICER collaboration belong to the basic set:
\begin{itemize}
    \item[{(II)}] $M = 1.44_{-0.14}^{+0.15}~M_{\odot}$ and $R = 13.02_{-1.06}^{+1.24}~\mathrm{km}$ for PSR J0030+0451 by~\citet{Miller:2019},
    \item[{(III)}] $M = 2.08_{-0.07}^{+0.07}~M_{\odot}$ and $R = 12.92_{-1.13}^{+2.09}~\mathrm{km}$ for  PSR J0740+6620 by~\citet{Dittmann:ApJ:2024}, which is a recent refinement of the previous NICER radius measurement by~\citet{Miller:2021}, and
    \item[{(IV)}]  $M = 1.418_{-0.037}^{+0.037}~M_{\odot}$ and $R = 11.36_{-0.63}^{+0.95}~\mathrm{km}$ for  PSR J0437-4715 by~\citet{Choudhury:2024}. 
\end{itemize}
These measurements (I) - (IV) form the basic data set for the present Bayesian inference.
The tidal deformability constraint from the binary neutron star merger GW170817 is also an important element of the Bayesian analysis
\begin{itemize}
    \item[{(V)}] $\Lambda_{1.4}=190^{+390}_{-120}$ for GW170817 by~\citet{LIGOScientific:2018cki}.
\end{itemize}
However, since we want to study the effect of its inclusion on the results of the Bayesian analysis, we display it separately.

Additionally, there are two extreme results that belong to an extended data set:
\begin{itemize}
    \item[{(VI)}] $M = 2.35_{-0.17}^{+0.17}~M_{\odot}$ for the ``black widow'' (BW) pulsar PSR J0952-0607 by~\citet{Romani:2022jhd}, and
    \item[{(VII)}] $M = 0.77_{-0.17}^{+0.20}~M_{\odot}$ and $R = 10.04_{-0.78}^{+0.86}~\mathrm{km}$ for HESS J1731-347 by~\citet{Doroshenko2022-uq} for which we use the shorthand notation ``HESS''. 
\end{itemize}
The additional constraints that were derived from observations of the binary neutron star merger GW170817 by \citet{Bauswein:2017vtn} and by \citet{Annala:2017llu} provide exclusion regions shown by red rectangular bars in the M-R diagram. 
They are, however, not taken into account into the Bayesian analysis of the present work.
The NICER mass-radius determinations for PSR J0740+6620 by~\citet{Riley:2021pdl} and for 
PSR J0030+0451 by~\citet{Raaijmakers:2019qny} as well as the mass measurement for PSR J0740+6620 by~\citet{Fonseca:2021wxt} which updated the earlier one by \citet{NANOGrav:2019jur} are also not considered in the present Bayesian analysis.


\begin{figure}[!htb]
    \centering
    \includegraphics[width=0.75\textwidth]{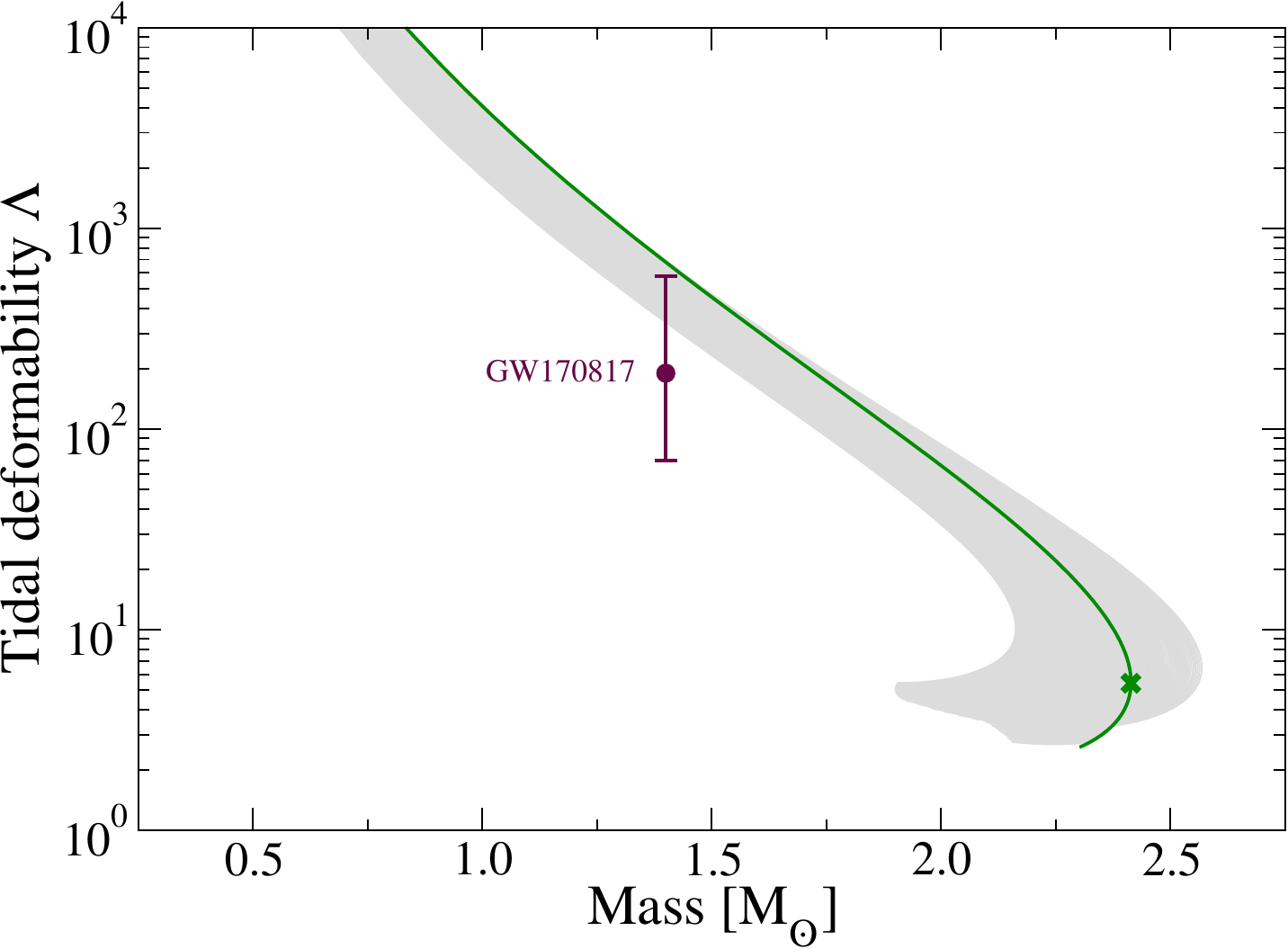}    
    \caption{Tidal deformability vs. mass 
 for the set of hybrid EOS introduced in this work. The green line highlights the result for the hadronic DD2 EOS and the green cross indicates the maximum mass configuration of that sequence.
 The {purple} data point with error bars corresponds to the gravitational wave analysis of the LIGO Virgo Collaboration for the inspiral phase of the binary neutron star merger GW170817 \cite{LIGOScientific:2018cki}.
    }
    \label{fig:LM}
\end{figure}

In Fig. \ref{fig:LM} we show the tidal deformability as a function of the neutron star mass for the hadronic DD2 reference EOS (green solid line) and the hybrid star sequences (grey lines) of the present study. The data point with error bars corresponds to the gravitational wave analysis of the LIGO Virgo Collaboration for the inspiral phase of the binary neutron star merger GW170817 \cite{LIGOScientific:2018cki}. The comparison indicates that hybrid star sequences with an early onset of quark deconfinement at masses well below $1.4~M_\odot$ lead to a softening of the EOS at the onset energy density which entails a reduction of the value of the tidal deformability at that mass and thus leads to a fulfillment of the tidal deformability constraint while the DD2 EOS was slightly too stiff to match that constraint.

\section{Bayesian analysis and results for constraining the EOS}
\label{sec:BA}

Let us describe the Bayesian framework used to infer the posterior distribution of the parameters \((\eta_D, \eta_V)\) defining EOS, given observational data \(\mathcal{D}\).
Bayes' theorem provides the posterior distribution of the parameters $(\eta_D, \eta_V)$,
\begin{equation}
    p((\eta_D,\eta_V) | \mathcal{D}) = \frac{p(\mathcal{D} | (\eta_D,\eta_V)) \, p(\eta_D,\eta_V)}{p(\mathcal{D})},
\end{equation}
where $p((\eta_D, \eta_V) | \mathcal{D})$ is the posterior distribution, $p(\mathcal{D} | (\eta_D, \eta_V))$ is the full likelihood, $p(\eta_D, \eta_V)$ is the prior distribution, $p(\mathcal{D})$ is the evidence. The prior distribution is taken to be equiprobable
\begin{equation}
    p(\eta_D, \eta_V)=\dfrac{1}{|{(\eta_D, \eta_V)}|}=\dfrac{1}{N}~.
\end{equation}
The full likelihood is obtained by production of the likelihoods for all independent observations \(D_\alpha\)
\begin{equation}
    p(\mathcal{D} | (\eta_D,\eta_V)) = \prod_{\alpha}^N p(D_{\alpha} | (\eta_D,\eta_V)).
\end{equation}
The likelihood associated with the lower limit of the maximum mass is modeled using a normal cumulative distribution function \(F_{\mathcal{N}}\),
\begin{equation}
    p(D_{M_{\max}^{(i)}} | (\eta_D,\eta_V)) = F_{\mathcal{N}}(M_{\max}(\eta_D,\eta_V); \mu_M^{(i)},\sigma_M^{(i)}),
\end{equation}
where \(\mu_M^{(i)}\) and \(\sigma_M^{(i)}\) are the mean and standard deviation of the constraint are taken form \cite{Antoniadis:Science:2013} and \cite{Romani:2022jhd}.

The likelihood for mass-radius constraints is computed via integration over the central density $\varepsilon_c$ with the appropriate probability density function
\begin{equation}
    p(D_{{MR}^{(i)}} | (\eta_D,\eta_V)) = \int_{\varepsilon_{c}^{\min}}^{\varepsilon_{c}^{\max}(\eta_D,\eta_V)} f^{(i)}_{MR}\left(M(\varepsilon_c;\eta_D,\eta_V), R(M)\right) pr(\eta_D,\eta_V) \, d\varepsilon_{c}.
\end{equation}
The probability density functions $f^{(i)}_{MR}$ have been constructed using Kernel Density Estimation (KDE)~\cite{Chacon2020-ql} based on data obtained from the Zenodo repository for MR constraints: PSR~J0030+0451~\cite{miller:2019:3473466}, PSR~J0740+6620~\cite{dittmann:2024:10215108}, and PSR~J0437–4715~\cite{choudhury:2024:13766753}, and HESS~J1731-347~\cite{doroshenko:2023:8232233}.
For gravitational wave (GW) data, the likelihood is expressed similarly as an integral over \(\varepsilon_c\)
\begin{equation}
    p(D_{GW} | (\eta_D,\eta_V)) = \int_{\varepsilon_{c}^{\min}}^{\varepsilon_{c}^{\max}(\eta_D,\eta_V)} f_{GW}\left(\Lambda_1(\varepsilon_c;\eta_D,\eta_V), \Lambda_2(\Lambda_1)\right) pr(\eta_D,\eta_V) \, d\varepsilon_{c}.
\end{equation}
Here, $f_{GW}$ is built using the data for GW170817 available from~\cite{ligoLIGOP1800115v12GW170817}.
The evidence \(p(\mathcal{D})\) is obtained by marginalizing over all possible values of \((\eta_D, \eta_V)\):
\begin{equation}
    p(\mathcal{D}) = \sum_{(\eta_D,\eta_V)} p(\mathcal{D} | (\eta_D,\eta_V)) \, p(\eta_D,\eta_V).
\end{equation}

\begin{figure}[!htb]
    \centering
    \includegraphics[width=
    \linewidth]{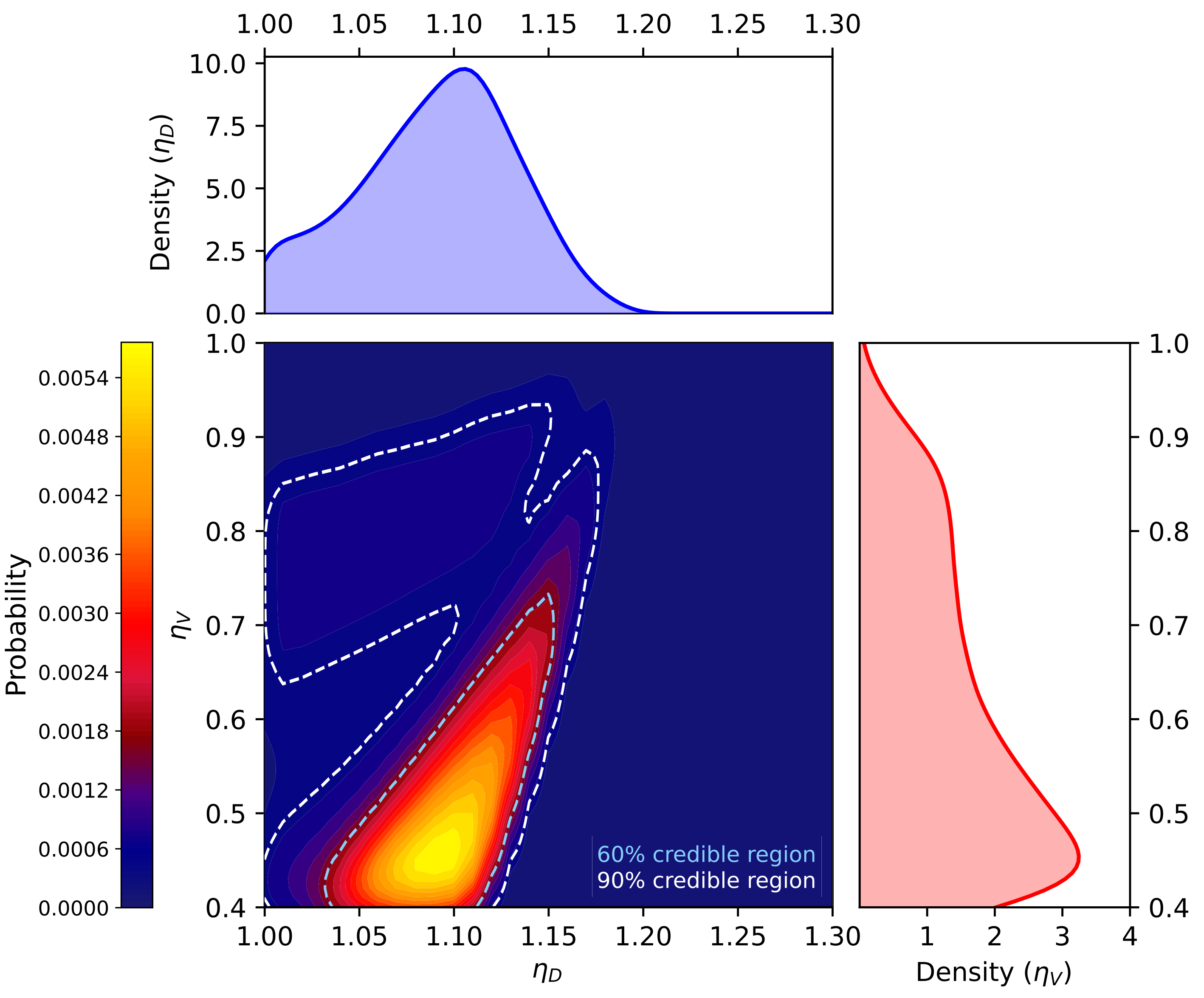}
   \caption{Bayesian analysis of the most likely parameter values in the $\eta_V$--$\eta_D$ plane under the modern observational constraints for masses, radii and the tidal deformability of neutron stars. }
    \label{fig:BA}
\end{figure}

The result of the Bayesian analysis is shown in Fig. \ref{fig:BA}.
We observe that the 60\% credible region forms a triangular shaped area closely resembling the nearly cigar-shaped region reported in previous works (e.g., in Refs. \cite{Baym:2019iky}, \cite{Gartlein:2023vif} and \cite{Gartlein:2024cbj}), although those results were not derived using a Bayesian analysis. 
The most favorable parameter set can be read off the peak positions for the marginalized probability distributions in the $\eta_V$- and $\eta_D$- directions, respectively. These peak positions are very close to the best choice values $\eta_V=0.5$ and $\eta_D=1.1$
that were found heuristically in \cite{Contrera:2022tqh}, where the present class of hybrid EOS was introduced.

\begin{figure}[!htb]
    \centering
    \includegraphics[width=0.85\linewidth]{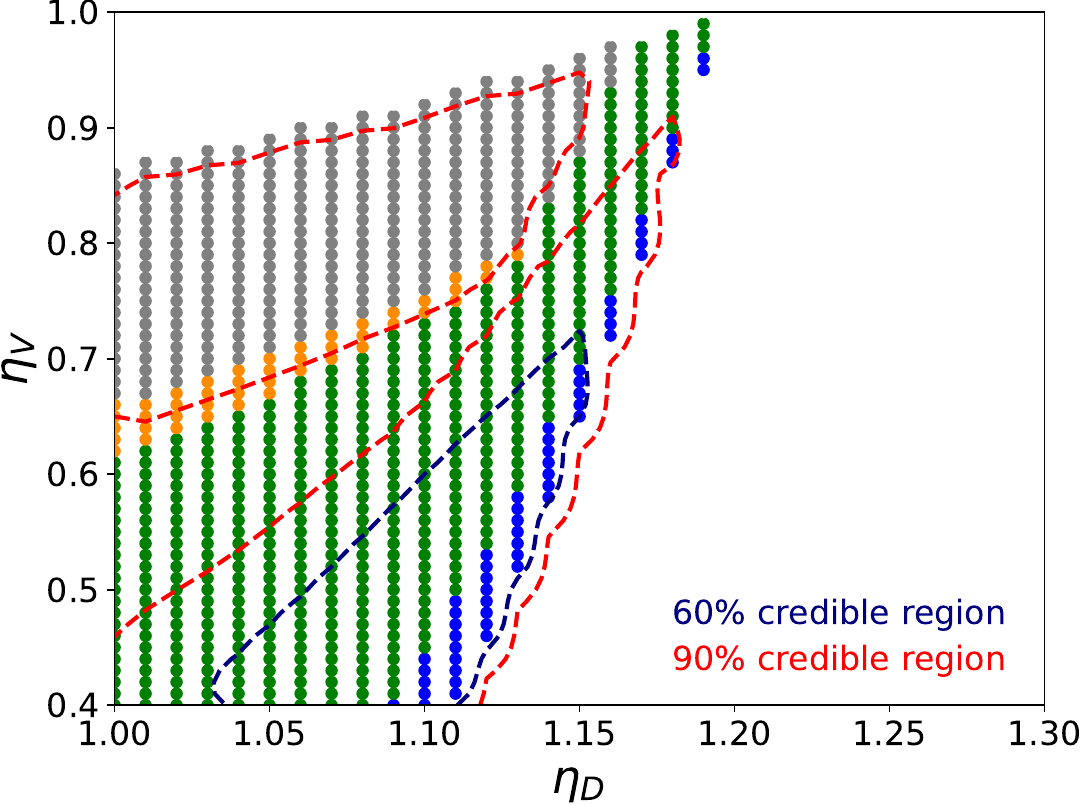}
    \caption{Phase transition categories in the $\eta_V$--$\eta_D$ diagram, overlaid with the 60\% (blue dashed line) and 90\% (red dashed lines) credible regions of the Bayesian analysis. For details, see text.}
    \label{fig:PhT-BA}
\end{figure}

In Fig. \ref{fig:PhT-BA} we show the different cases of phase transitions, with the following color coding:
\begin{itemize}
    \item[]{- White:} No Maxwell transition,
    \item[]{- Gray:} Maxwell transition occurs after the DD2 instability value (no stable hybrid stars),
    \item[]{- Orange:} Maxwell transition starts before the DD2 instability value, with a jump to the DD2 instability (no stable hybrid stars),
    \item[]{- Green:} Maxwell transition ends before the DD2 instability value (stable hybrid stars exist) 
    \item[]{- Blue:} Maxwell transition ends before the DD2 instability value and satisfies the Seidov criterion\footnote{The Seidov criterion \cite{1971SvA15347S} states that for phase transitions with $\Delta\varepsilon \ge (\varepsilon_{\rm onset}+3p_{\rm onset})/2$ a gravitational instability of the compact star configuration is to be expected.}, leading to extremely light twin stars. 
\end{itemize}
Overlaid on this classification are the credible regions of the Bayesian analysis with confidence levels of 60\% (blue dotted lines) and 90\% (red dotted lines). 
One can read off from this figure not only the favorable parameter region and deduce the corresponding set of hybrid EOS, but also whether a hybrid star sequence with a color superconducting quark core or even a mass twin solution is favorable. 

\begin{figure}[htbp]
    \centering
    \begin{tabular}{cc}
        \begin{adjustbox}{valign=t}
        \subfloat[Basic set of constraints.]{%
            \includegraphics[width=0.475\textwidth]{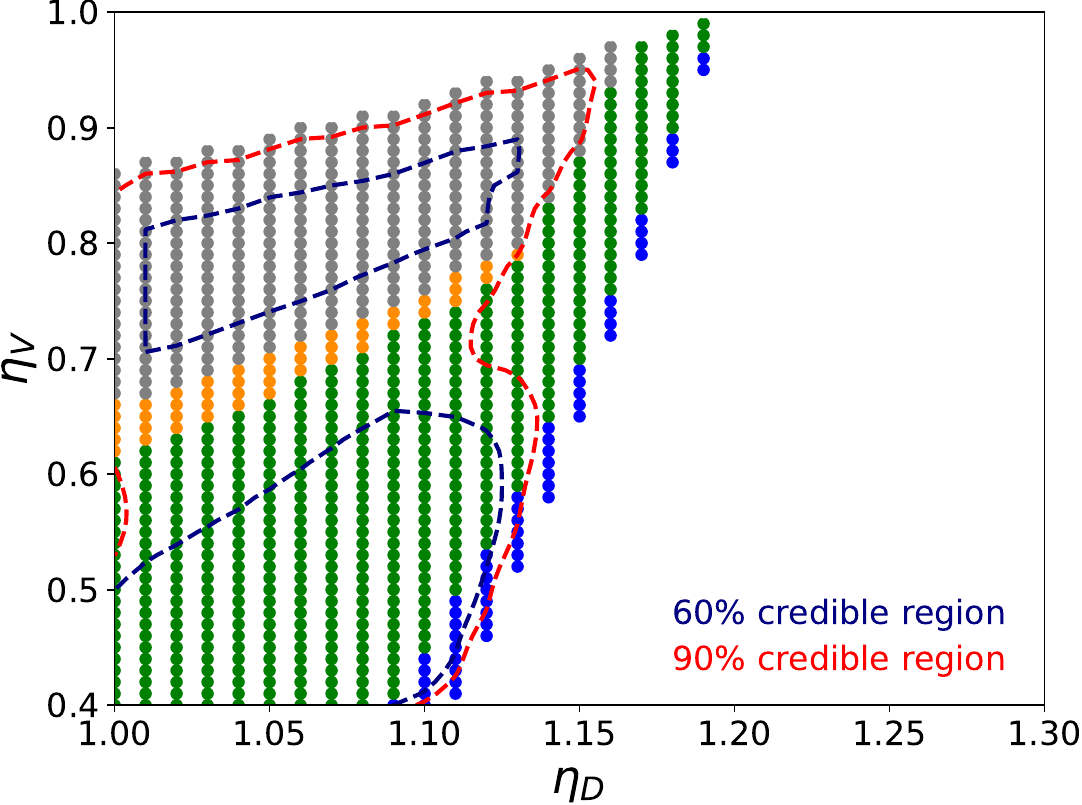}
            \label{fig:PhT-BA-zoo:basic}
        }
        \end{adjustbox} &
        \begin{adjustbox}{valign=t}
        \subfloat[Full set of constraints.]{%
            \includegraphics[width=0.475\textwidth]{FigBA/Hybrid_EoS_DD2_QM3DFF_PhT_prop_BA_Dittmann_full.pdf}
            \label{fig:PhT-BA-zoo:full}
        }
        \end{adjustbox} \\
        \begin{adjustbox}{valign=t}
        \subfloat[Basic set plus HESS.]{%
            \includegraphics[width=0.475\textwidth]{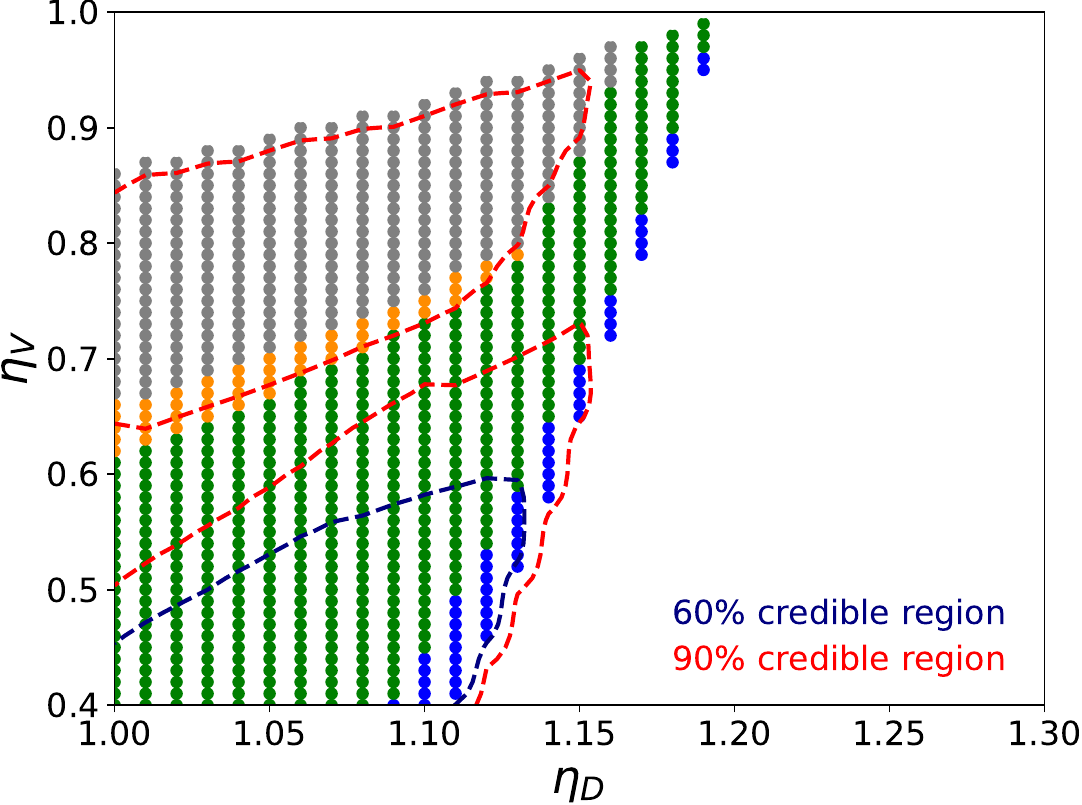}
            \label{fig:PhT-BA-zoo:basic-HESS}
        }
        \end{adjustbox} &
        \begin{adjustbox}{valign=t}
        \subfloat[Basic set plus BW and HESS.]{%
            \includegraphics[width=0.475\textwidth]{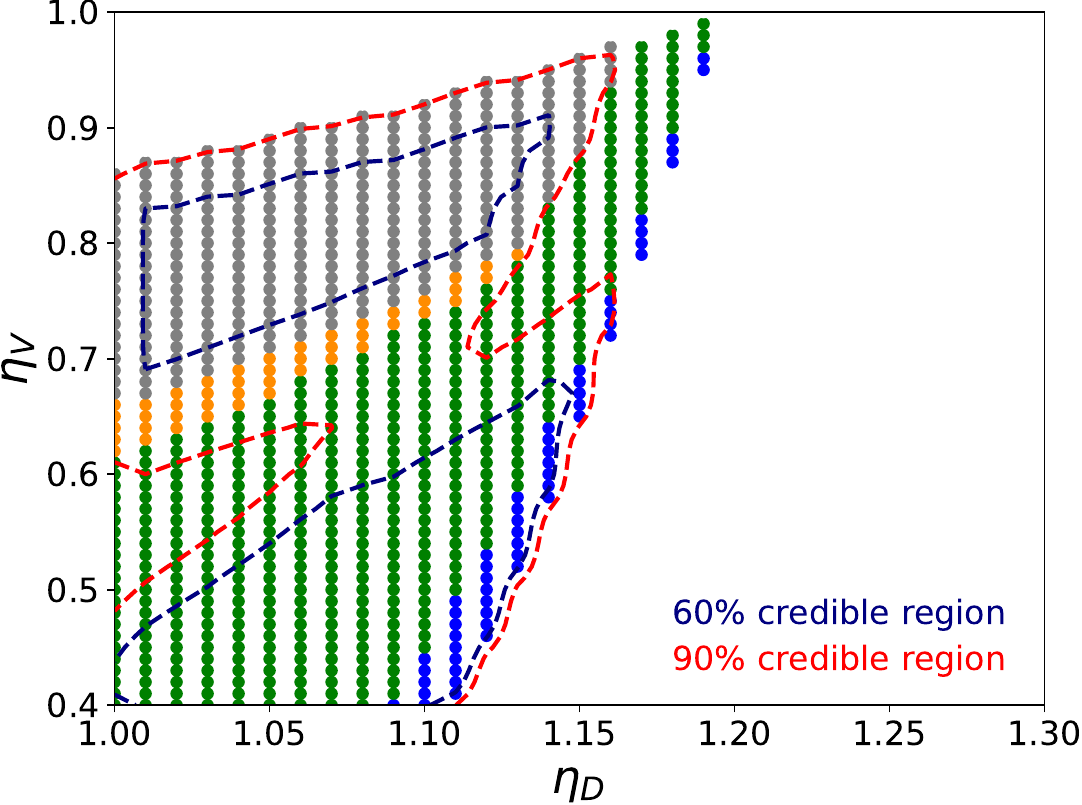}
            \label{fig:PhT-BA-zoo:basic-BW-HESS}
        }
        \end{adjustbox} \\
        \begin{adjustbox}{valign=t}
        \subfloat[Basic set plus GW170817.]{%
            \includegraphics[width=0.475\textwidth]{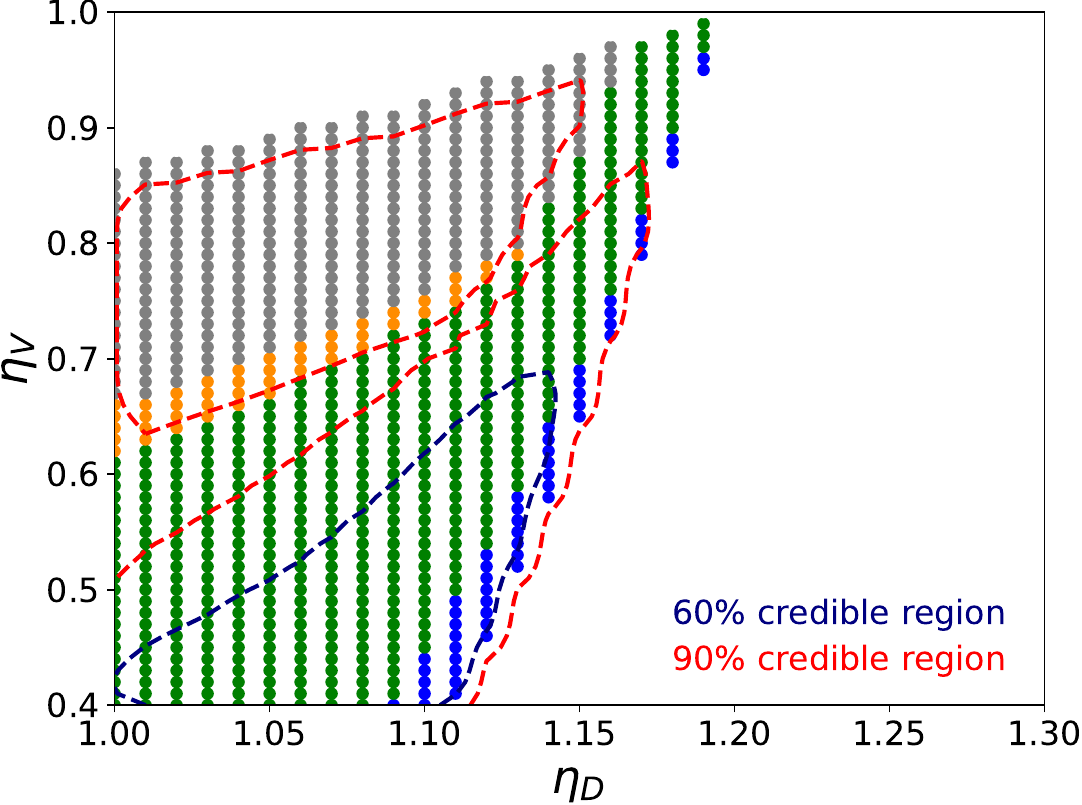}
            \label{fig:PhT-BA-zoo:basic-GW}
        }
        \end{adjustbox} &
        \begin{adjustbox}{valign=t}
        \subfloat[Basic set plus BW and GW170817.]{%
            \includegraphics[width=0.475\textwidth]{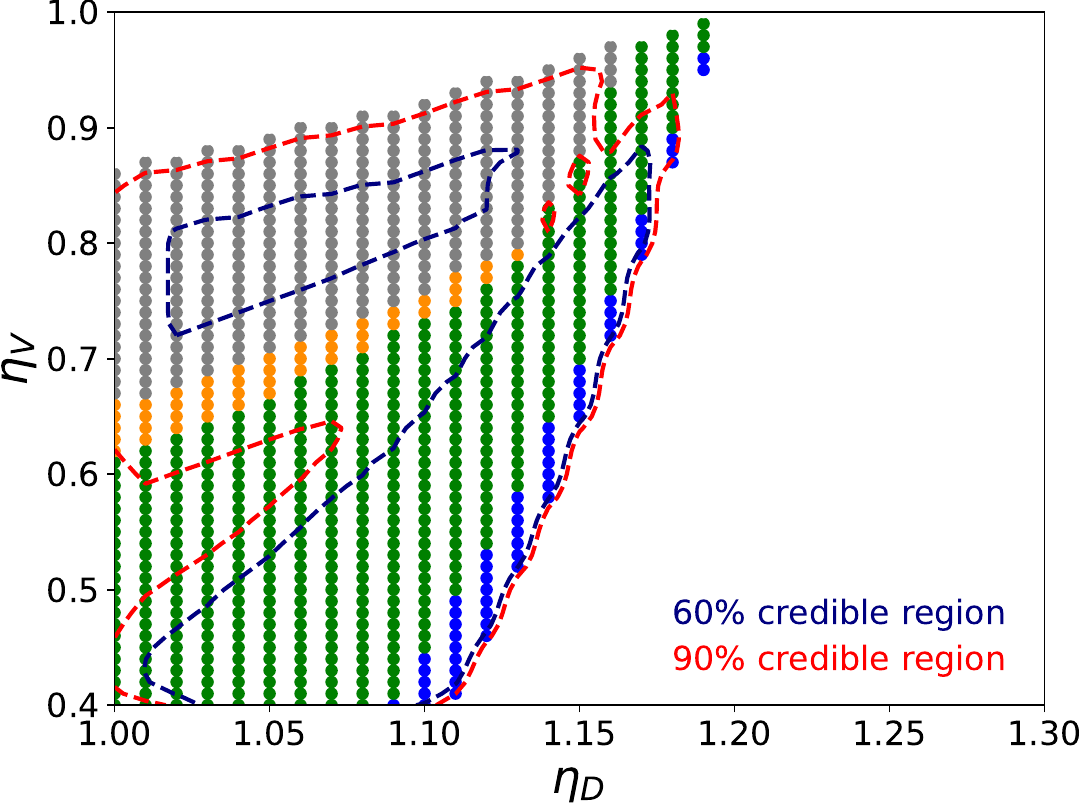}
            \label{fig:PhT-BA-zoo:basic-BW-GW}
        }
        \end{adjustbox} \\
        \begin{adjustbox}{valign=t}
        \subfloat[Basic set plus GW170817 and HESS.]{%
            \includegraphics[width=0.475\textwidth]{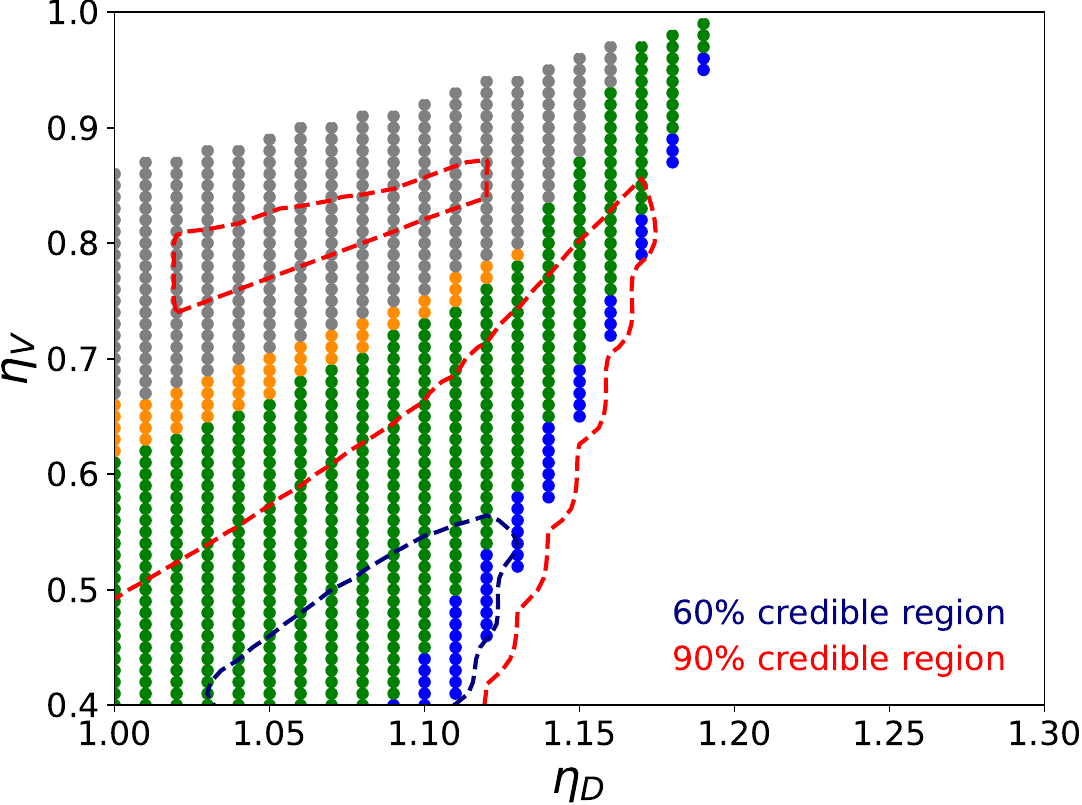}
            \label{fig:PhT-BA-zoo:basic-GW-HESS}
        }
        \end{adjustbox} &
        \begin{adjustbox}{valign=t}
        \subfloat[Basic set plus BW.]{%
            \includegraphics[width=0.475\textwidth]{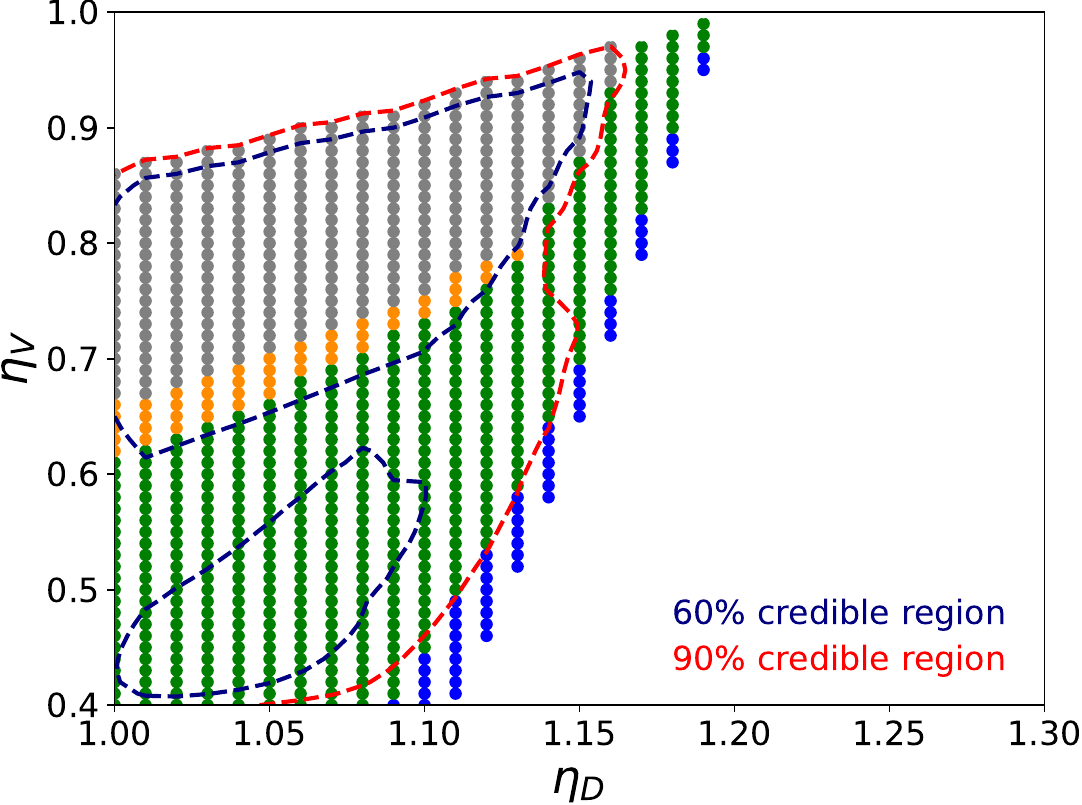}
            \label{fig:PhT-BA-zoo:basic-BW}
        }
        \end{adjustbox}
    \end{tabular}
    \caption{Same as Fig. \ref{fig:PhT-BA} for different sets of constraints. The constraints (I), (II), (III), and (IV) constitute the basic set shown in panel (\textbf{a}), while the full set is shown in panel (\textbf{b}), identical to \mbox{Figure \ref{fig:PhT-BA}.} Additional constraints (V), (VI), and (VII) are denoted as GW170817, BW, and HESS, respectively, and analysed in different combinations with the basic set in panels (\textbf{c}--\textbf{h}).}
    \label{fig:PhT-BA-zoo}
\end{figure}

In Fig. \ref{fig:PhT-BA-zoo} we show results of investigating the role that the choice of certain sets of observables has for the topology of favorable regions in the parameter plane as a result of the BA. On eight panels (a) - (h) we show the 60\% and 90\% confidence regions overlaid to the four types of hybrid star sequences. Starting from panel (a) with the basic set of constraints that excludes the tidal deformability measurement of GW170817, we see two separate regions encircled with the blue dashed lines for the 60\% credibility. One lies entirely inside the grey domain which belongs to purely hadronic configurations where no stable hybrid stars exist, the other one is entirely in the green domain corresponding to standard hybrid stars with not too early onset of deconfinement ($\eta_D\lesssim 1.1$) and no twin stars. Since the stiff DD2 hadronic baseline EOS with a maximum mass of $M=2.414\, M_\odot $ (case A in Tab. \ref{tab:ns_props} reaches well beyond the high-mass constraints (I) and (III), the hybrid star sequences safely fulfill these constraints with not too large vector meson couplings ($\eta_V\lesssim 0.6$).

When the lower limit on the maximum mass is increased to by adding the black widow pulsar constraint (VI) to the basic set, then the 60\% credibility region at large $\eta_V$ is enlarged and comprises all purely hadronic stars as well as hybrid stars with not too early onset of deconfinement ($\eta_D\lesssim \eta_{D, \rm max} - 0.05$ for any $\eta_V$) see panel (h). 

When we add constraints to the basic set which require compactness at low mass (GW170817 and/or HESS), we observe that the probability for the purely hadronic EOS is suppressed and the second 60\% region in the grey domain is absent on panels (c), (e) and (g). This holds also for the full set of constraints shown on panel (b).  

Adding the tidal deformability constraint (V) from GW170817 to the basic set requires the onset masses to be lowered and therefore this drags the high-$\eta_D$ border of the 60\% credibility region at lower-$\eta_V$ to maximally admissible $\eta_D$ values, where even mass twin sequences are possible, see panel (e). This shape is only stretched towards the corner of highest couplings in the parameter plane when additionally the BW constraint (VI) is included, see panel (f). 
Since panel (f) is very similar to panel (b) for the full set of constraints, one can conclude that out of all additional constraints (V) - (VII), the latter is dominated by the tidal deformability constraint.

\begin{figure}[!htb]
    \centering
    \includegraphics[width=0.85\linewidth]{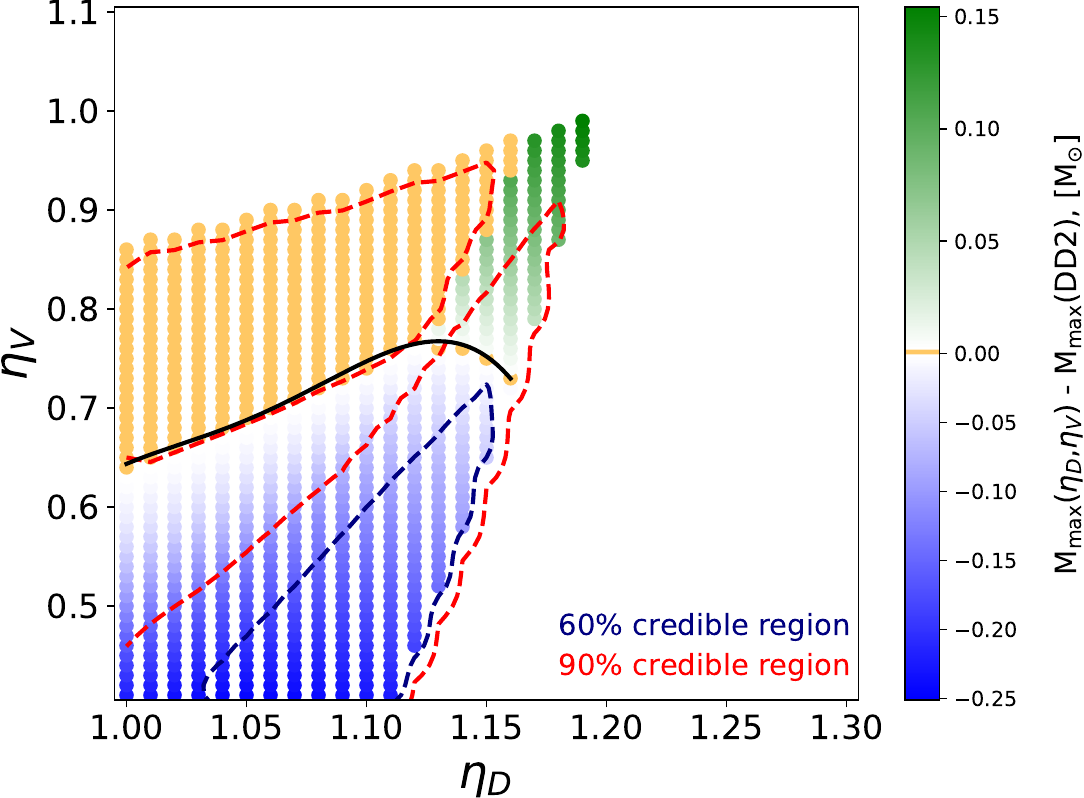}
    \caption{Color-bubble diagram for the difference in maximum mass between hybrid stars and the DD2 hadronic reference EOS with a maximum mass of 2.4 $M_\odot$ (the black solid neutrality line)
    in the $\eta_V$-$\eta_D$ plane of EOS parameters overlaid to the Bayesian likelihood contours for the 60\% and 90\% credibility regions.}
    \label{fig:BA-Mmax}
\end{figure}

\begin{figure}[!htb]
    \centering
    \includegraphics[width=0.85\linewidth]{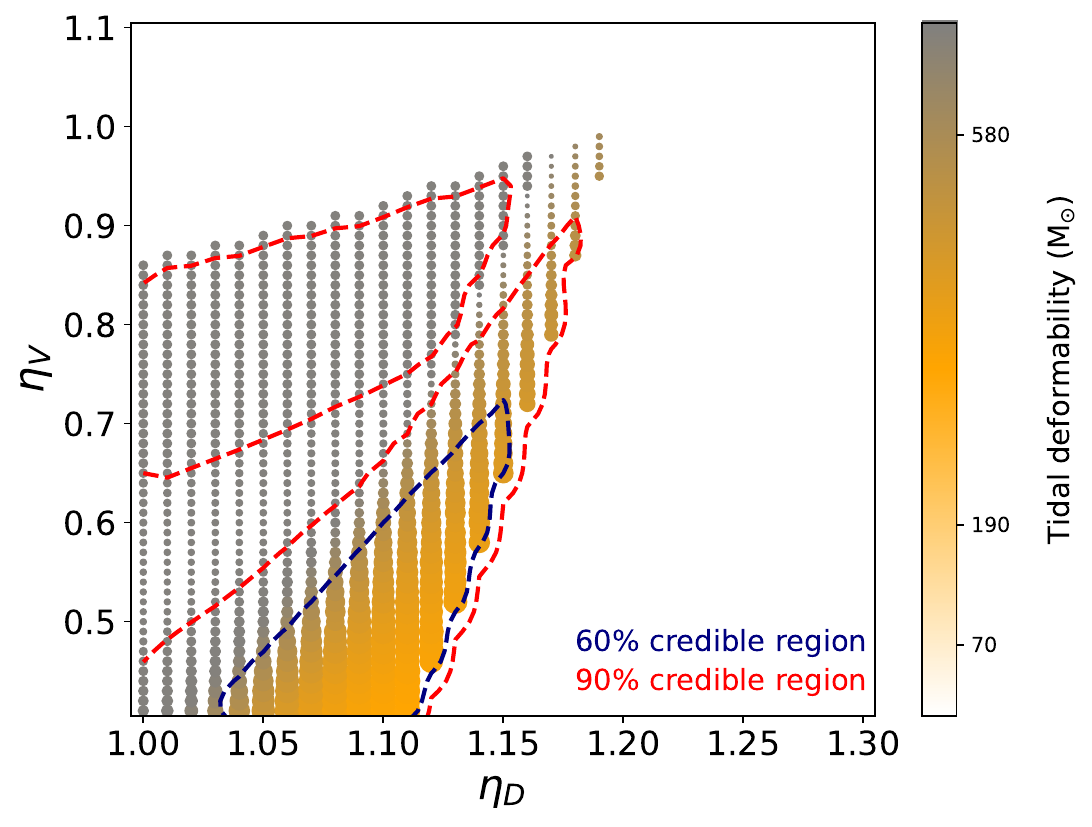}
    \caption{Color-bubble diagram for the tidal deformability in the $\eta_V$-$\eta_D$ plane of EOS parameters overlaid to the Bayesian likelihood contours for the 60\% and 90\% credibility regions. On the color legend bar the constraint (V) from the LVC measurement of the tidal deformability for GW170817 is indicated: $70 < \Lambda_{1.4}< 580$.}
    \label{fig:BA-Deformabilit}
\end{figure}

In Figs. \ref{fig:BA-Mmax} and \ref{fig:BA-Deformabilit}, we display observables in the plane of the EOS parameters $\eta_V$ and $\eta_D$ as color-bubble diagrams overlaid to the Bayesian likelihood contours for the 60\% and 90\% credibility regions. 
In Figs. \ref{fig:BA-Mmax} we show the difference in maximum mass between hybrid stars and the DD2 hadronic reference EOS with a maximum mass of 2.4 $M_\odot$ (the black solid neutrality line). The yellow bubble region indicates that no stable hybrid stars are possible there. 

We want to mention that in \cite{Baym:2019iky}, the maximum possible mass was at $2.35~M_\odot$  and in \cite{Gartlein:2023vif,Gartlein:2024cbj}  at $2.42~M_\odot$. 
Thus the corresponding parameter sets of the latter works would lie in the corner of the green region, between the 60\% and 90\% confidence regions, while the present class of hybrid EOS has the paramneter sets with the largest maximum masses at the very corner of the green region, outside even the 90\% confidence region. 

The reason for this region of highest-mass hybrid EOS to be disfavored lies in the fact that they predict hybrid star sequences that are not sufficiently compact in the mass region of $1.4~M_\odot$, where the tidal deformability was measured for the binary neutron star merger GW170817 and requires radii in the range $R_{1.4}=10.94 - 12.61$ km at 90\% confidence \cite{Dietrich:2020efo}.  
In Figs. \ref{fig:BA-Deformabilit} we show the tidal deformability constraint: $70 < \Lambda_{1.4}<580$ and can observe that the 60\% credible region (blue dashed line) encloses the pairs of quark matter parameter values for which the tidal deformability constraint is fulfilled.

\begin{figure}[!htb]
    \centering
    \includegraphics[width=0.9\linewidth]{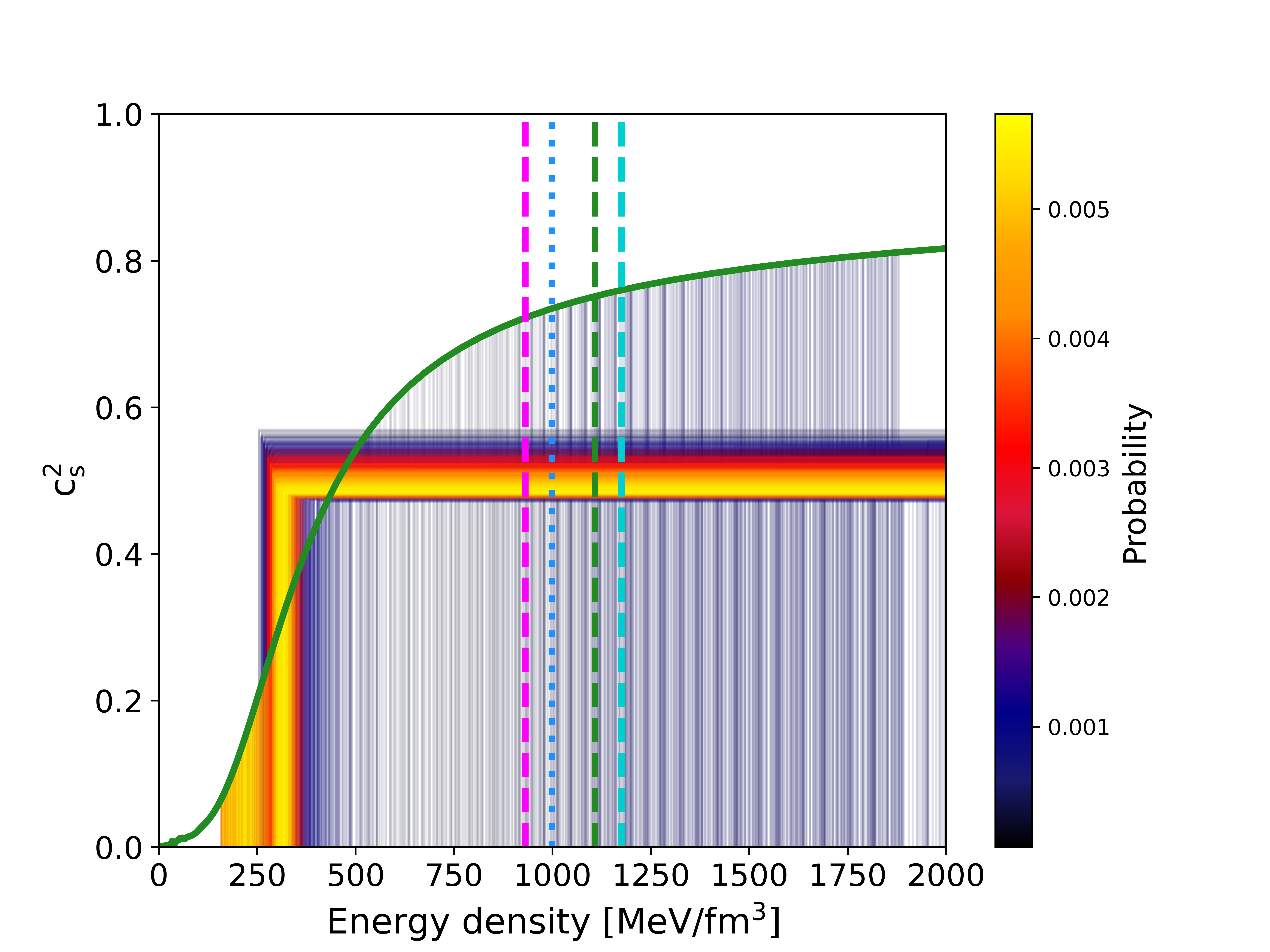}
    \caption{Squared speed of sound as a function of the energy density for the hadronic DD2 reference EOS (green solid line), the hybrid star EOS with color superconducting 3DFF quark matter fitted to $c_S^2=const$ (colored lines) with the color code shown in the legend corresponding to the posterior probability after Bayesian analysis.
In the first-order phase transition region holds $c_S^2=0$. The onset (end) of the deconfinement phase transition is marked by vertical lines connecting the bottom of the graph at $c_S^2=0$ with the green DD2 line (the horizontal $c_S^2=const$ lines).  
The bold vertical lines indicate the central energy densities of the maximum mass configurations for case A: the purely hadronic EOS (green dashed line), case C: the hybrid EOS with the highest 
posterior probability (blue dashed line) and case D: the hybrid EOS with the highest maximum mass (magenta dashed line) as well as of the onset mass configurations for case B: the hybrid EOS with the highest onset density (blue dotted line). 
}
    \label{fig:cs2-BA}
\end{figure}

In Fig. \ref{fig:cs2-BA} we show the squared speed of sound as a function of the energy density for the hadronic DD2 reference EOS (green solid line) and the hybrid star EOS with a Maxwell-constructed first-order phase transition to color superconducting 3DFF quark matter fitted to $c_S^2=const$ with a color code of the lines representing the probability after Bayesian analysis as shown in the legend. In the first-order phase transition region holds $c_S^2=0$. 
We would like to comment on the fact that the values of the squared speed of sound fitted to the EOS of the color superconducting nonlocal chiral quark matter model \cite{Contrera:2022tqh} lie in a narrow range around $c_s^2\sim 0.5$. This deviation from the conformal limit value  $c_s^2=1/3$ 
can be traced even analytically to the appearance of color superconductivity in quark matter.
Based on the effective EOS for color superconducting quark matter by Alford, Braby, Paris and Reddy \cite{Alford:2006vz} (see also \cite{Zhang:2020jmb}), one can obtain the formula
$c_s^2=(1+\zeta)/(3+\zeta)$, where $\zeta\propto \Delta^2/\mu^2$ with $\Delta$ being the diquark pairing gap, the order parameter signaling color superconductivity \citet{Blaschke:2022egm}. 
For $\zeta=1$ follows  $c_s^2 = 1/2$ and for normal quark matter ($\Delta=0$) the conformal limit holds. For a detailed discussion of the behavior of the squared speed of sound in the color superconducting nonlocal chiral quark matter and the approach to the conformal limit, see \cite{Ivanytskyi:2024zip}.

The maximum energy density that can be reached in the center of a neutron star made of DD2 matter for its maximum mass configuration is $\varepsilon_c^{\rm max}=1108$ MeV/fm$^3$ (vertical green dashed line, case A) and reduces to $\varepsilon_c^{\rm max}=998.5$ MeV/fm$^3$ (vertical blue dotted line, case B) when a deconfinement transition is possible but does not lead to stable hybrid stars.
The central energy density of the maximum mass configuration for the hybrid EOS with the highest 
posterior probability (light blue dashed line, case C) is $\varepsilon_c^{\rm max}=1175$ MeV/fm$^3$
while that of the hybrid EOS with highest maximum mass (magenta dashed line, case~D) is 
$\varepsilon_c^{\rm max}=931$ MeV/fm$^3$. The neutron star properties for cases A--D are summarized in~Table~\ref{tab:ns_props}.

\begin{table}[!h]
\caption{Neutron star properties for different special hybrid EOS cases: A -- purely hadronic DD2; B -- highest onset energy density with a stable hybrid star branch; C -- highest posterior probability; D -- highest maximum mass; E -- highest maximum mass in the 90\% credibility area. For details, see text.}
\label{tab:ns_props}
\hspace*{-5mm} 
\begin{tabular}{|c|c|c|c|c|c|c|c|c|c|}
\hline
&($\eta_D$, $\eta_V$) & M$_{\max}$ & R$_{\max}$ & $\varepsilon_{c,{\rm max}}$ & $\mu_{c,{\rm max}}$ & M$_{\mathrm{onset}}$ & R$_{\mathrm{onset}}$ & $\varepsilon_{c,{\rm onset}}$ & $\mu_{c,{\rm onset}}$ \\ 
& -- & M$_{\odot}$ & km & $\frac{\mathrm{MeV}}{\mathrm{fm}^3}$ & MeV & M$_{\odot}$ & km & $\frac{\mathrm{MeV}}{\mathrm{fm}^3}$ & MeV \\ \hline \hline
A&DD2 & 2.414 & 11.84 & 1107.88 & 1909.04& -- & -- & -- & -- \\ \hline
B&(1.12, 0.76) & 2.411 & 12.01 & 1116.29 & 1820.92 & 2.410 & 12.03 & 998.54 & 1810.84 \\ \hline
C&(1.08, 0.44) & 2.195 & 11.42 & 1175.19 & 1709.54 & 0.630 & 12.92 & 234.47 & 1017.95 \\ \hline
D&(1.19, 0.99) & 2.568 & 12.54 & 931.18 & 1776.96 & 0.323 & 13.84 & 188.38 & 983.82 \\ \hline
E&(1.18, 0.91) & 2.521 & 12.38 & 956.42 & 1767.16 & 0.330 & 13.78 & 196.10 & 984.87 \\ \hline
\end{tabular}
\end{table}

\begin{figure}[!htb]
    \centering
    \includegraphics[width=0.85\linewidth]{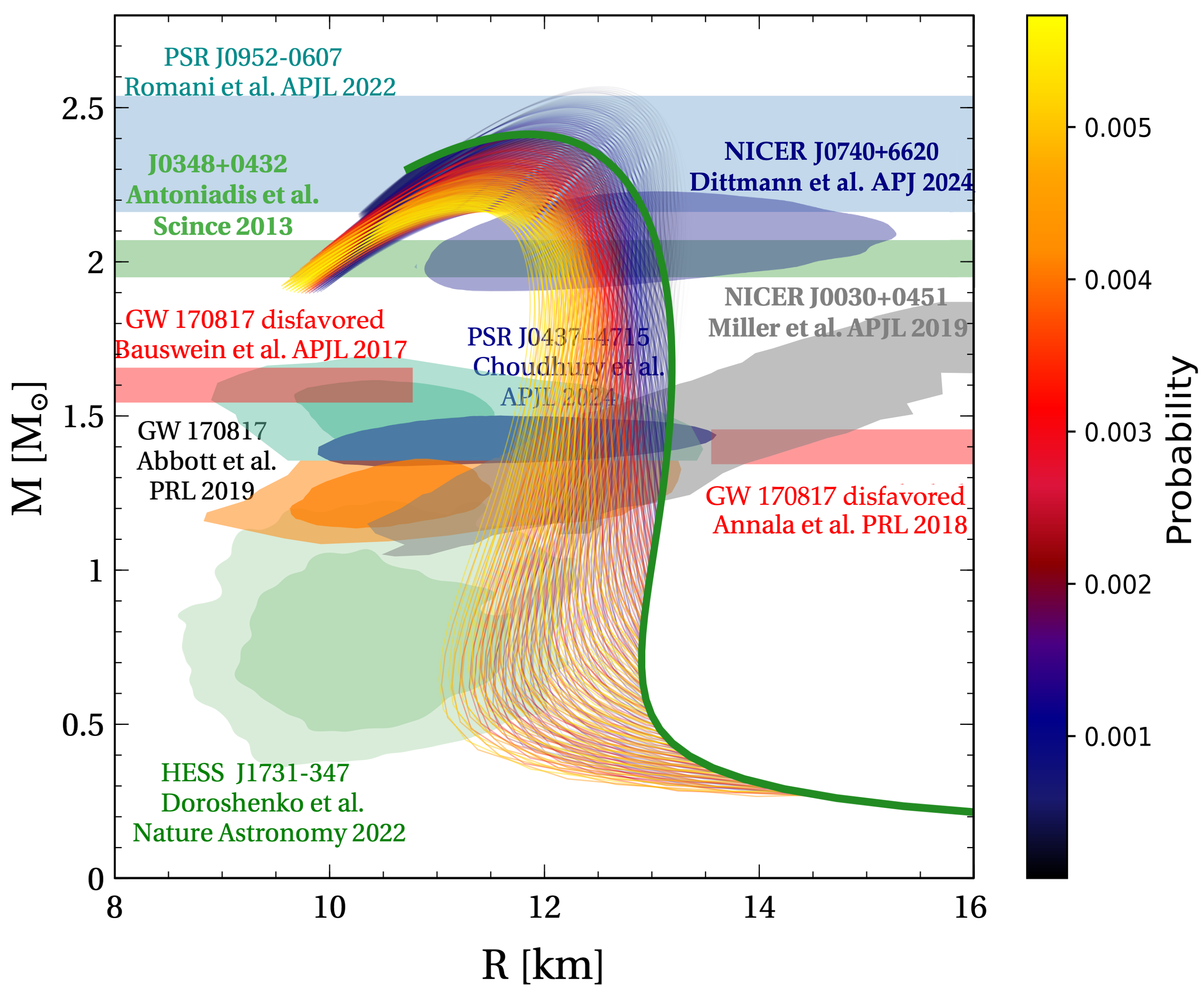}
    \caption{Mass-radius diagram showing the hadronic DD2 reference EOS (green solid line) and the hybrid star sequences corresponding to the hybrid EOS Maxwell-constructed from it with the CSS fitted form of the nonlocal 3DFF color superconducting quark matter EOS ((colored lines) with the color code shown in the legend corresponding to the posterior probability after Bayesian analysis with the modern mass-radius constraints (colored and labelled regions). }
    \label{fig:MR-BA}
\end{figure}

The final result of this Bayesian analysis study for the class of hybrid EOS with the hadronic DD2 RDF reference EOS and the nonlocal 3DFF color superconducting quark matter EOS fitted to the CSS form of EOS is shown in Fig. \ref{fig:MR-BA}. 
We note that the EOS parametrisations favored by the Bayesian analysis with the displayed set of observational constraints listed in section \ref{sec:BA} provide evidence for an early onset of deconfinement below $0.7~M_\odot$
and maximum masses not exceeding $2.25~M_\odot$.
In the region of observed pulsar masses, $M\sim 1.1 - 2.1 ~M_\odot$, the radii are approximately mass independent in the range of $R\sim 12~$ km. 
This feature has been discussed in Ref. \cite{Ferreira:2024hxc} as an indicator for maximum masses in the range $M_{\rm max}\sim 2.3 - 2.4 ~M_\odot$ within a model-independent study.
This is consistent with our result shown in Fig. \ref{fig:MR-BA}.

One may raise the question whether this study can be considered realistic because it neglects the strangeness degree of freedom in the hadronic as well as in the quark matter phase. 
We expect that the main conclusions of our study, the favorability of an early onset of deconfinement in the mass range 0.5-0.7 $M_\odot$ and that maximum masses of the most favorable hybrid star sequences do not exceed 2.25 $M_\odot$, would not be affected by the inclusion of strangeness in the hadronic and quark matter phases.
Since the onset density for hyperons in the hadronic phase is well above the onset of deconfinement it would not affect the early onset of deconfinement. What concerns the maximum mass of the hybrid neutron stars with a strange quark matter core, we would expect that a readjustment of the vector meson and diquark coupling strengths would be necessary. But within a Bayesian analysis, they should result in basically the same favorable range of maximum masses as in the 2-flavor quark matter case.

\section{Conclusions}

In this work, we have performed for the first time a physics-informed Bayesian analysis of the EOS constraints from modern observational data for masses, radii and the tidal deformability of pulsars within the class of hybrid EOS with a stiff nuclear matter phase, color superconducting (two-flavor) quark matter and a Maxwell construction for the phase transition. 
The color superconducting phase of quark matter along with repulsive vector meson interactions is essential for ensuring that the hybrid compact star model matches observational constraints because it allows a sufficiently low threshold density for the onset of deconfinement and a maximum mass of at least 2 $M_\odot$.
This scheme supersedes an earlier physics-informed Bayesian analysis study \cite{Ayriyan:2021prr} that was performed within the alternative scheme that uses a soft nuclear matter EOS together with a two-zone interpolation scheme for the transition to color superconducting quark matter.
While the latter scheme provided hybrid EOS models in the entire rectangular-shaped section of the 
two-dimensional EOS parameter space spanned by the vector-meson ($\eta_V$) and diquark ($\eta_D $) couplings, the impossibility to perform a Maxwell construction in the former scheme removed the two corners at high-$\eta_V$--low-$\eta_D$  and high-$\eta_D$--low-$\eta_V$ in the present work. 

Inside the remaining region highlighted by colored dots in Fig. \ref{fig:PhT-BA}, where the Maxwell construction of a phase transition is possible, 
the following generic structures can be delineated in the $\eta_V-\eta_D $ plane and are highlighted on Fig. \ref{fig:PhT-BA} by dots in different colors:
\begin{itemize}
    \item the rightmost line of proportionality $\eta_V \propto \eta_D $ with the early onset of deconfinement, eventually accompanied with the mass twin phenomenon (blue dots),
    \item the region of stable hybrid stars (green dots),
    \item the upper right corner of this region, where the hybrid stars with highest maximum mass are found which are eventually disfavored because of missing compactness. In the present work, this corner is at $\eta_V= 0.99,~ \eta_D = 1.19$, corresponding to $M_{\rm max} = 2.57~M_\odot$ at $R_{\rm max}=12.4$ km.
    \item the region parameter region, where the onset of deconfinement is before the maximum central energy density of purely hadronic neutron stars, 
    $\varepsilon_{c,~\rm onset}< \varepsilon_{c,~\rm max}$ (orange dots), where there are no stable hybrid stars but due to the presence of a deconfinement transition in the EOS the maximum mass is lowered relative to the purely hadronic case, 
    \item the region, where the onset of deconfinement occurs after the maximum central energy density of purely hadronic neutron stars, 
    $\varepsilon_{c,~\rm onset}> \varepsilon_{c,~\rm max}$ (grey dots).
\end{itemize}

The main result of the present Bayesian analysis are constrained regions of 60\% and 90\% credibility in the $\eta_V-\eta_D $ plane which allow to further constrain the triangular-shaped region that was found earlier in heuristic studies without employing Bayesian analysis methods \cite{Klahn:2013kga,Baym:2019iky,Gartlein:2023vif,Gartlein:2024cbj}.
We found that EOS in the upper right corner of that triangle which correspond to hybrid stars with maximum masses above the purely hadronic maximum mass of $2.41~M_\odot$ are incompatible with the 
60\% credibility region of the Bayesian analysis! 
This appears to be a consequence of the tidal deformability constraint that disfavors these high-maximum mass hybrid star sequences because of their insufficient compactness. 
We demonstrated how the choice of a set of mass-radius constraints influences the shape of the credibility regions in the plane of EOS parameters. 
Small neutron star radii at or below typical binary radio pulsar masses (e.g., $R_{1.4} \lesssim 12$ km) as for HESS and GW170817 require an early onset of deconfinement as induced by large values of $\eta_D \gtrsim 1.1$ (color superconductivity) while the vector coupling responsible for the stiffness of the EOS and large maximum masses can stay at moderate values of $\eta_V \lesssim 0.5$.
Including the back widow pulsar PSR J0952-0607 into the set of constraints would require EOS with a maximum mass of at least $2.18~M_\odot$ (the 1$\sigma$ level). In such a case, the Bayesian analysis results in strong vector couplings $\eta_V \gtrsim 0.5$ for hybrid star sequences. 
We note that for our choice of a stiff hadronic baseline EOS such as DD2, the maximum mass constraint is always fulfilled already without a deconfinement phase transition. This is reflected also in the Bayesian analysis result which exhibits the parameter region with no stable hybrid stars as favorable at 60\% confidence level.  

The next step in the development of this physics-informed Bayesian analysis method is to guarantee the fulfilment of the requirement that the high-density asymptotics of the hybrid EOS family should be compatible with the perturbative QCD benchmark at densities $n\ge 40~n_0$, see \cite{Komoltsev:2021jzg,Gorda:2022jvk, PhysRevLett.127.162003} by using a quark matter EOS model that obeys the conformal limit and matches the pQCD EOS.  
A promising and straightforward strategy to achieve this goal is the relaxation of the locality assumption for the vector mean field. This has recently been demonstrated by Ivanytskyi \cite{Ivanytskyi:2024zip}. 
Furthermore, the strangeness degree of freedom should be included in both the hadronic and quark matter phases of the study.
Work in this direction is in preparation. 

\vspace{6pt}

\authorcontributions{Conceptualization and methodology, A.A. and D.B.; software, A.A., J.P.C. and G.A.C.; validation and formal analysis, A.A., G.A.C. and A.G.G.; investigation, A.A.; data curation, A.A.; writing---original draft preparation, A.A. and D.B.; writing---review and editing, A.A., D.B. and A.G.G.; visualization, A.A.; supervision and project administration, D.B.; funding acquisition, all authors. 
All authors have read and agreed to the published version of the manuscript.}

\funding{A.A. and D.B. were supported by NCN under grant No. 2021/43/P/ST2/03319.
A.G.G., G.A.C., and J.P.C. would like to acknowledge
CONICET, ANPCyT and UNLP (Argentina) for financial support under grants No. PIP 2022-2024 GI - 11220210100150CO, PICT19-00792, PICT22-03-00799 and X960, respectively.}

\dataavailability{Data will be made available by the authors upon reasonable request.}

\acknowledgments{We thank Oleksii Ivanytskyi for providing us with the template of the present observational mass and radius constraints for pulsars that we used in preparing Figures 
\ref{fig:MR} and \ref{fig:MR-BA}.}

\conflictsofinterest{The authors declare no conflicts of interest.}




%

\appendixtitles{no} 

\begin{adjustwidth}{-\extralength}{0cm}

\reftitle{References}




\PublishersNote{}
\end{adjustwidth}
\end{document}